\documentclass[aps,pra,twocolumn,amsmath,amssymb]{revtex4-2}

\usepackage{graphicx}
\usepackage{color}
\usepackage[hidelinks]{hyperref}
\usepackage{orcidlink}

\hypersetup{colorlinks=true, linkcolor=black, citecolor=black, urlcolor=blue}

\begin{document}
\author{Karol Gietka\,\orcidlink{0000-0001-7700-3208}}
\email[]{karol.gietka@uibk.ac.at}

\affiliation{Institut f\"ur Theoretische Physik, Universit\"at Innsbruck\\Technikerstra{\ss}e\,21a, A-6020 Innsbruck, Austria} 

\title{Vacuum Rabi splitting as a manifestation of virtual two-mode squeezing: \\ Extracting the squeezing parameters from frequency shifts}


\begin{abstract}
Vacuum Rabi splitting relies on symmetrical splitting of the common resonance frequency of atoms and the cavity in which the atoms reside. In this work, we argue that vacuum Rabi splitting is a manifestation of virtual light-matter two-mode squeezing. We establish a connection between squeezing parameters of virtual modes and frequency shifts of the physical modes. To this end, we use the mapping between the Dicke model and two interacting harmonic oscillators, which we analyze in the framework of bare and physical modes. Finally, we suggest that such virtual squeezing of quantum fields might also play a role in quantum field theories.
\end{abstract}
\date{\today}
\maketitle


\section{Introduction}
Light-matter interaction is a fundamental concept in quantum science and technology~\cite{QT2003milburn,Acín_2018} as well as foundational physics~\cite{scully1999quantum,weiner2003light,walmsley2015scienceQO}. The quantum electrodynamics describes interaction between light and matter in a fully quantum and relativistic manner~\cite{greiner2013quantum}. It gives extremely accurate predictions of various phenomena, in particular, Lamb shift of the hydrogen energy levels~\cite{lambshif1947} or the anomalous magnetic moment of the electron~\cite{electronmoment1948schwinger}, and is often said to be the most tested and most precise theory in physics. The key element in quantum electrodynamics are virtual excitations~\cite{cao2004conceptual} which modify properties of bare field modes and give rise to physical modes whose properties can be measured in the experiments~\cite{carusotto2005vacuum,loudon2000quantum}. The effect of virtual excitations is especially significant in the ultrastrong-coupling regime in cavity quantum electrodynamics where only one mode of the electromagnetic field is isolated from the entire spectrum~\cite{solano2019rmpUSC,nori2019reviewUSC}. These systems typically consist of a qubit (an artificial atom) coupled to a resonator in which the effective light-matter coupling can be tuned or switched on and off which opens vast possibilities for exploring quantum vacuum fluctuations and the relation between physical and bare particles. Unfortunately, despite several theoretical proposals, the experimental detection of virtual excitations still awaits demonstration~\cite{carusotto2007vacuumradiation,boas2008rabivacuum,savasta2013virtualtoreal,nori2016electrolumin,nori2017virtualpressure,paraoanu2024detectvirtual}.

\begin{figure}[htb!]
    \centering
    \includegraphics[width=0.49\textwidth]{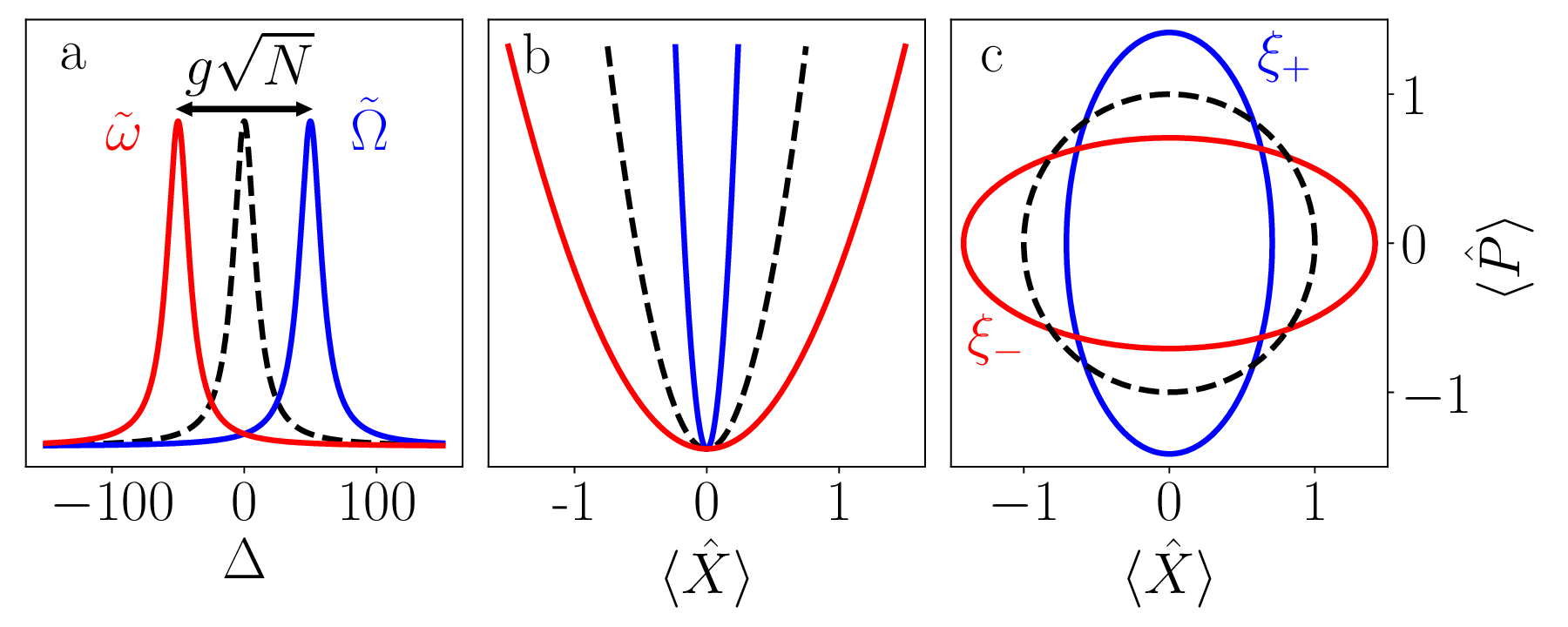}
    \caption{Scheme elucidating the connection between frequency shifts and virtual squeezing, $\tilde \omega = \omega e^{2\xi_-}$ and $\tilde \Omega = \omega e^{2\xi_+}$. (a) shows vacuum Rabi splitting of the cavity resonance frequency (dashed black line) into two peaks with increased frequency (blue line) and decreased frequency (red line). In the limit of $g/g_c \ll 1$, the frequency shifts are linear. (b) electromagnetic field with different frequencies can be understood as harmonic oscillators with different frequencies. (c) ground states of harmonic oscillators with different frequencies are squeezed with respect to each other which can be depicted in the (quadrature) phase space~\cite{squeezing1987frequency,CFLo_1990_squeezingfrequency}. This squeezing is, however, virtual and its observation remains so far elusive in light-matter systems.}
    \label{fig:scheme}
\end{figure}

In many situations entering the ultrastrong-coupling regime is challenging. One of the most prominent example is coupling the atomic transition to the optical frequency of a cavity through an electric dipole moment~\cite{helmut2013rmp}. In its full form, such coupling is typically described by the quantum Rabi model~\cite{rabimodel1937}. In optical domain, however, the light-matter coupling is negligible with respect to the bare resonance frequencies of the atom and the cavity, therefore the effect of virtual excitations seemingly can be neglected. Consequently, the quantum Rabi model can be treated with the rotating wave approximation under which the counter-rotating terms get eliminated leading to the paradigmatic Jaynes-Cummings model~\cite{JC2021larson}. One of the many predictions of this influential model is the so-called vacuum Rabi splitting (also known as normal mode splitting) which manifests in splitting the cavity resonance frequency symmetrically into two peaks~\cite{vrs2983eberly,vrs2984agarwal,vrs1992kimble}, depicted in Fig.~\ref{fig:scheme}(a). The separation between the peaks can be increased $\sqrt{N}$ times by coupling $N$ atoms to the cavity leading to the Tavis-Cummings model. Such enhanced vacuum Rabi splitting can be used for non-destructive measurements of atoms in the cavity~\cite{thompson2011vrsSS} and thus for creating spin-squeezing~\cite{MA201189}. Note that vacuum Rabi splitting can be also observed in solid state micro cavities~\cite{vrs1992ssmc,ciurri2009nature}.

In this manuscript, we argue that vacuum Rabi splitting is a manifestation of virtual entanglement in the form of two-mode squeezing and that the frequency shifts are related to squeezing parameters. To this end, we consider the paradigmatic Dicke model under the Holstein-Primakoff approximation which we analyze in the framework of bare and physical modes. We generalize the vacuum Rabi splitting to the Dicke model, analytically establish connection between virtual squeezing and frequency shifts, numerically calculate the maximal number of virtual excitations that can be generated in a finite system composed of $N$ atoms, and briefly discuss why observing virtual excitations directly is so far elusive.


\section{virtual squeezing in the Dicke Model}
The Dicke Hamiltonian is a paradigmatic model in quantum physics that was historically used to describe a collection of two level atoms coupled to a single mode of radiation~\cite{dickemodel2011garraway}. The Hamiltonian of the Dicke model reads ($\hbar \equiv 1$)
\begin{align}\label{eq:dickeH}
    \hat H = \omega \hat a^\dagger \hat a + \Omega \hat S_z  + \frac{g}{\sqrt{N}}(\hat a^\dagger + \hat a)\hat S_x,
\end{align}
where $\hat a^\dagger$ creates an excitation of the electromagnetic field (photon) with frequency $\omega$, $\hat S_i$ with $i=x,y,z$ describe the two-level atoms with atomic transition energy $\Omega$, and $g$ is the light-matter coupling strength. In the above equation, the insertion of $N$ ensures the critical coupling does not depend on the system size. The Dicke model, exhibits the so-called superradiant phase transition~\cite{HEPP1973360} which has been a subject of intense theoretical and experimental investigations in recent decades~\cite{srpt1973hioe,wodkiewicz1975ptnogo,birula1979nogoSR,chaosDicke2003BrandesEmary,MASCHLER2007446,SRPT2011marquardt,feshke2012qptdicke,plenio2015qrm,hemmerich2015dynamic,Larson_2017,amr2018ionDM,nori2023srpt,kim2024observation,farokh2024dickemodels}. The superradiant phase transition is characterized by a macroscopic occupation of mode $\hat a$ and occurs once the coupling strength overcomes the critical coupling $g_c \equiv \sqrt{\omega \Omega}$ which however is forbidden in natural systems due to the Thomas-Reiche-Kuhn sum rule~\cite{wodkiewicz1975ptnogo,birula1979nogoSR}. Notably, the superradiant phase transition has been simulated in the rotating frame of driven-dissipative light-matter systems~\cite{esslinger2010srpt}, and very recently using artificial atoms coupled to a microwave cavity~\cite{nori2023srpt}. Superradiant phase transition can be also explored in systems where the electromagnetic radiation is replaced by a mechanical harmonic oscillator~\cite{amr2018ionDM} or magnons~\cite{kim2024observation}.

\begin{figure}[htb!]
    \centering
    \includegraphics[width=0.49\textwidth]{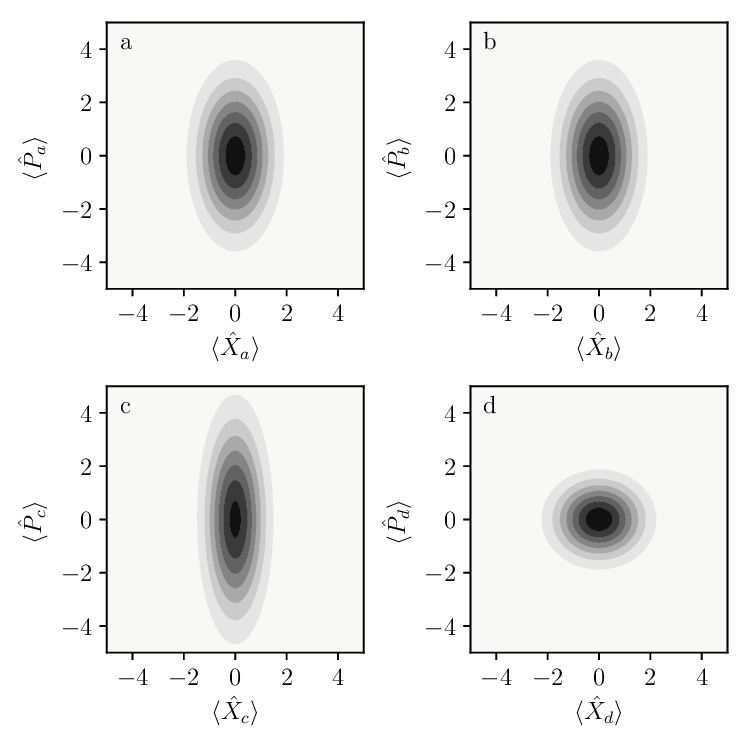}
    \caption{Squeezing of the ground state in the framework of virtual excitations. (a), (b), (c), and (d) depicts quadrature phase space (Husimi function) squeezing of mode $\hat a$, $\hat b$, $\hat c$, and $\hat d$, respectively. Interestingly, the squeezing of the bare modes $\hat a$ and $\hat b$ is the same, whereas the squeezing of the hybrid mode $\hat c$ is much more significant than squeezing of the hybrid mode $\hat d$. The physical modes $\hat e$ and $\hat f$ (not shown) are never squeezed. For the simulations we have set $g/g_c = 0.99$.}
    \label{fig:fig1}
\end{figure}

\begin{figure}[htb!]
    \centering
    \includegraphics[width=0.49\textwidth]{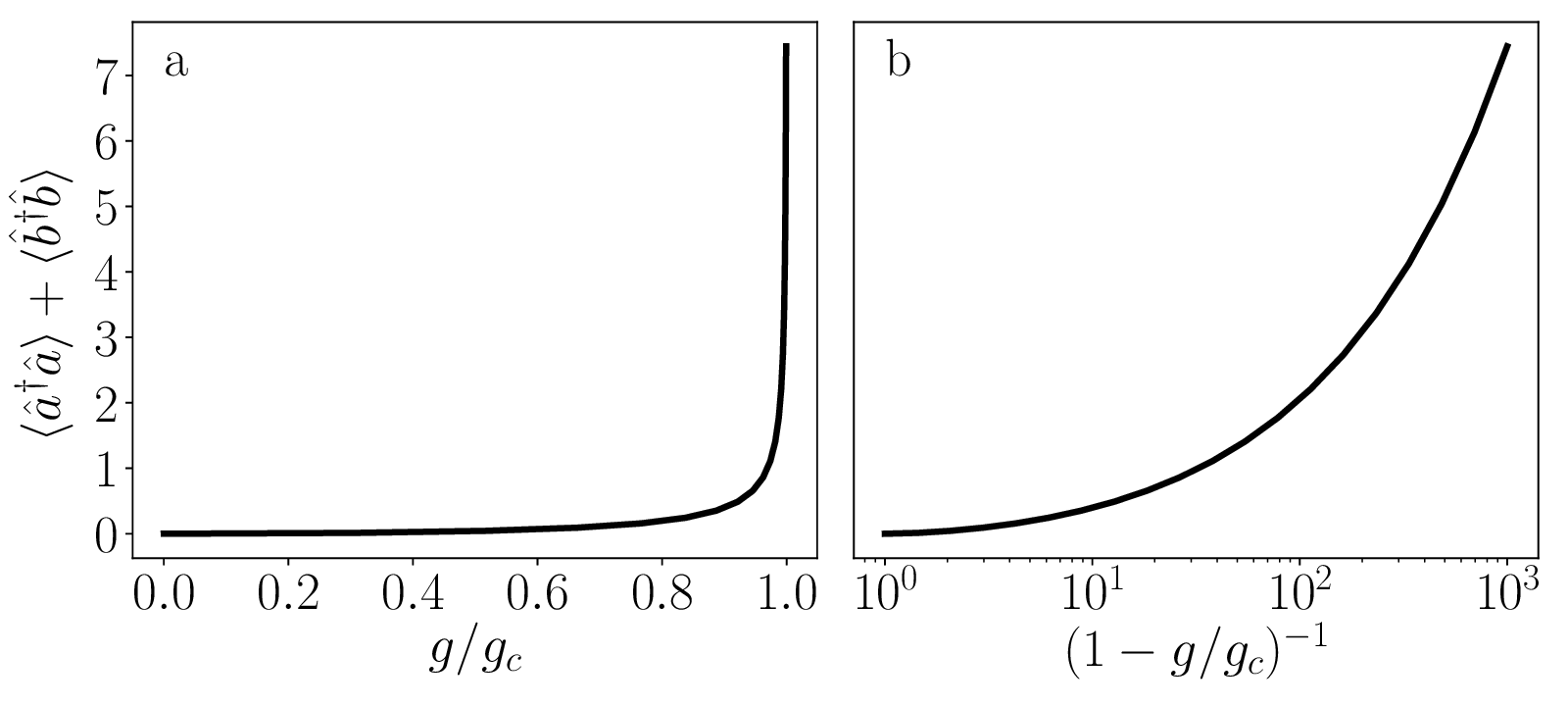}
    \caption{The number of virtual excitations as a function of $g/g_c$ is singular at the critical point. However, in order to generate a large amount of virtual excitations, the coupling $g$ has to be extremely close to the critical coupling rate. (a) depicts dependence on $g/g_c$, whereas (b) depicts dependence on $(1-g/g_c)^{-1}$ to elucidate singularity of the virtual excitations number at the critical point of the superradiant phase transition.}
    \label{fig:fig2}
\end{figure}

If the number of two-level atoms $N$ is large, it is convenient to apply the Holstein-Primakoff transformation~\cite{HP1940PR} and $N\rightarrow \infty$ approximation under which the Dicke Hamiltonian describes two interacting harmonic oscillators
\begin{align}\label{eq:twoH}
      \hat H = \omega \hat a^\dagger \hat a + \Omega \hat b^\dagger \hat b  + g(\hat a^\dagger + \hat a)(\hat b^\dagger + \hat b),
\end{align}
and can be treated analytically. In the following, we will consider a resonant case $\omega = \Omega$ which is particularly relevant for atoms coupled to optical cavities. Using the bare mode operators $\hat a$, $\hat a^\dagger$, $\hat b$, and $\hat b^\dagger$ to analyze the ground state of the Hamiltonian~\eqref{eq:twoH}, we find that the lowest energy state is an exotic two-mode squeezed vacuum (for typical two-mode squeezing see Refs~\cite{twomodeS1985theory,twomodeS1987exp}), presented in Fig.~\ref{fig:fig1}, which can be expressed using two squeeze operators with different but related squeezing parameters acting on the vacuum of modes $\hat a$ and $\hat b$  as
\begin{align}
   |\xi_-,\xi_+\rangle = e^{\frac{\xi_-}{2}(\hat c^2 - \hat c^{\dagger 2})} e^{\frac{\xi_+}{2}(\hat d^2 - \hat d^{\dagger 2})}|0_{a},0_{b}\rangle,
\end{align}
where $\hat c = (\hat a - \hat b)/\sqrt{2}$ and $\hat d = (\hat a + \hat b)/\sqrt{2}$ are the hybrid light-matter modes that are being squeezed and $\xi_- =\frac{1}{4}\log[1-g/g_c] $ and $\xi_+ =\frac{1}{4}\log[1+g/g_c]$ are the corresponding squeezing parameters. The exotic two-mode squeezing is particularly significant close to the critical point of the superradiant phase transition because the squeezing parameter $\xi_-$ becomes singular once $g=g_c$ (see Fig.~\ref{fig:fig2}). This is also related to closing the energy gap. Note that similar argument holds for approaching the critical point from the other side of the superradiant phase transition~\cite{chaosDicke2003BrandesEmary}.

\begin{figure}[htb!]
    \centering
    \includegraphics[width=0.49\textwidth]{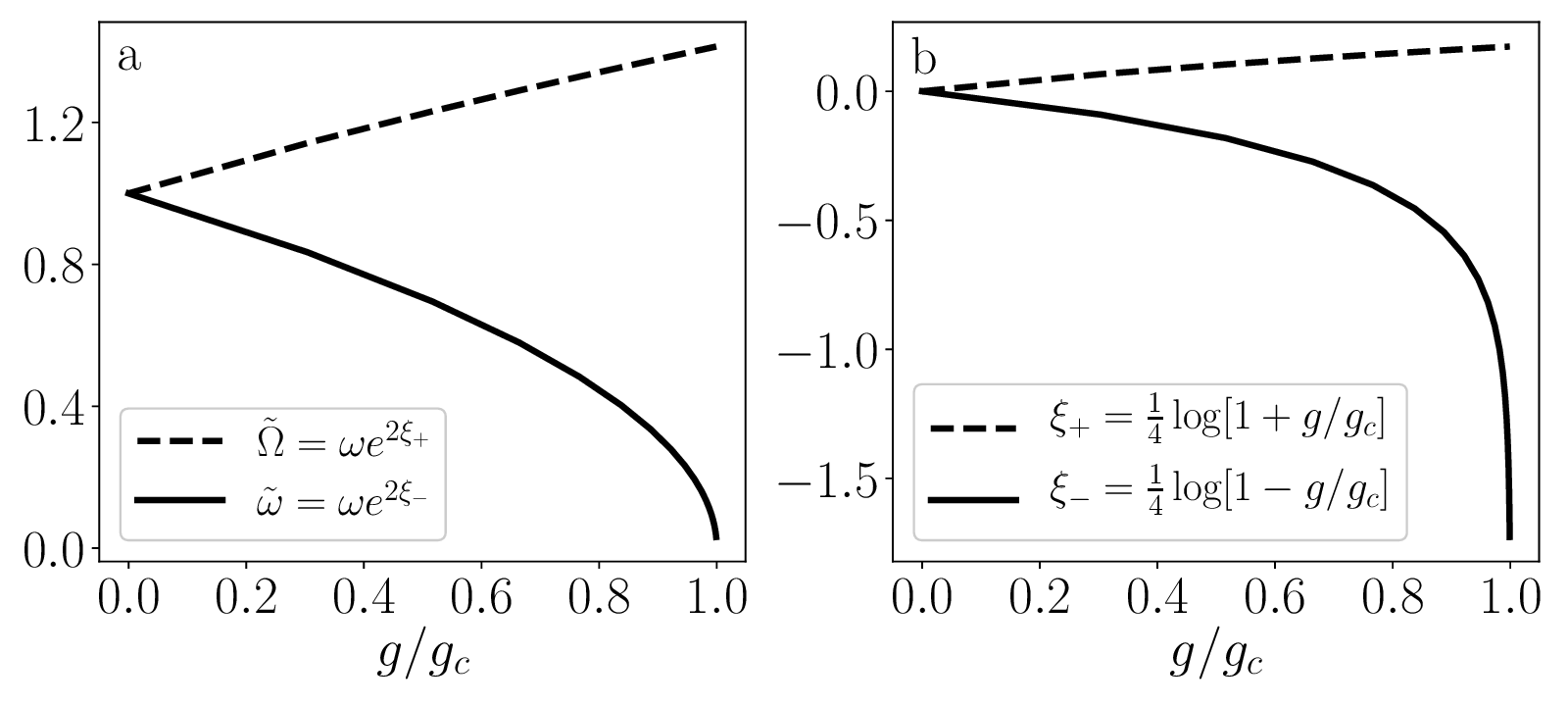}
    \caption{Relation between squeezing of the virtual excitations and frequency shifts of the physical modes. (a) depicts frequency shifts of two interacting harmonic oscillators. In the regime where $g/g_c\ll 1$ the frequency shifts are linear which is known as the vacuum Rabi splitting. (b) depicts the associated squeezing of the virtual excitations which can only be measured by suddenly turning off the light-matter coupling or by determining the frequency shifts.}
    \label{fig:fig3}
\end{figure}

Interpreting naively $\hat a$ and $\hat a^\dagger$ as operators annihilating and creating cavity photons, one would come to a conclusion that even before the superradiant threshold the cavity is filled with photons which could radiate through imperfect mirrors (see Fig.~\ref{fig:fig2}). This, however, cannot be the case because the ground state cannot lose energy by definition~\cite{carusotto2005vacuum,ciuti2006input,carusotto2007vacuumradiation}. Likewise, interpreting naively $\hat b$ and $\hat b^\dagger$ as the operators describing atomic modes, one would come to a conclusion that system could (spontaneously) emit radiation because the atoms are excited. Close to the critical point, the bare operators from Hamiltonian~\eqref{eq:twoH} do not correspond to measurable but virtual excitations which affect the properties of physical modes. The correct operators for the physical modes can be build out of the bare operators using the Bogoliubov transformation which diagonalizes the Hamiltonian~\cite{chaosDicke2003BrandesEmary} 
\begin{align}\label{eq:noentD}
    \hat H = \tilde \omega \hat e^\dagger \hat e + \tilde \Omega \hat f^\dagger \hat f,
\end{align}
where $\tilde \omega = \omega\sqrt{1-g/g_c} = \omega e^{2\xi_-}$ , $\tilde \Omega = \omega\sqrt{1+g/g_c} = \omega e^{2\xi_+}$ are the eigen frequencies of the system representing the physical modes (often called polaritons~\cite{Basov2021,novel2021QED}) and are related to squeezing parameters $\xi_-$ and $\xi_+$ (see Fig.~\ref{fig:fig3}). This means that the squeezing of the virtual excitations is represented by frequency shifts (squeezed and unsqueezed oscillators) of the physical modes. In other words, whenever $g\neq 0$ the physical modes are represented by 
\begin{align}
    \hat e =  e^{\frac{\xi_-}{2}(\hat c^{\dagger 2} - \hat c^{ 2})} \, \hat c  \, e^{\frac{\xi_-}{2}(\hat c^2 - \hat c^{\dagger 2})}, 
\end{align}
and
\begin{align}
\hat f =  e^{\frac{\xi_+}{2}(\hat d^{\dagger 2} - \hat d^{ 2})} \, \hat d \, e^{\frac{\xi_+}{2}(\hat d^2 - \hat d^{\dagger 2})}.
\end{align} 
This means that for $g \neq 0$, $\hat a$ and $\hat b$ are not the jump operators and one cannot drive the mode $\hat a$ or $\hat b$ either as they do not correspond to physical modes~\cite{blais2011dissipation,plenio2018dissprabi,Cattaneo_2019,solano2019rmpUSC,gietka2023uniqueSSqrm}. Moreover, it means that the entanglement of bare modes $\hat a$ and $\hat b$ cannot loose coherence once $g\neq 0$.

In order to convert virtual excitations into real ones, one would have to transform $\hat a$ and $\hat b$ back into operators describing the bare modes~\cite{carusotto2005vacuum,ciuti2006input,carusotto2007vacuumradiation}. This could be, in principle, done by instantaneously turning off the light-matter coupling. As a consequence of eliminating the coupling term from the Hamiltonian~\eqref{eq:twoH}, the energy stored in the squeezed vacuum could be released. Putting it differently, once the coupling is turned off, the ground state of a strongly coupled light-matter system close to the critical point becomes a highly excited and entangled state of the uncoupled system which means that it can lose energy. As a result, the cavity could emit light entangled with the atomic states (see also Appendix for a simpler case in the quantum Rabi model). 

Note, however, that for $g \ll g_c$, which is a typical condition in atom-light experiments~\cite{helmut2013rmp} where the coupling is several orders of magnitude lower than frequencies, the amount of entanglement between the bare modes is negligible and the bare modes $\hat a$ and $\hat b$ are almost identical to the physical modes $\hat e$ and $\hat f$. This is the essence of the rotating wave approximation as the frequency shifts are linear and squeezing is negligible (see Fig.~\ref{fig:fig3}). In the limit of $g\ll g_c$, the two frequencies of physical modes $\hat e$ and $\hat f$ can be approximated as $\tilde \omega \approx \omega - g/2$ and $\tilde \Omega \approx \omega + g/2$ which is exactly the symmetric vacuum Rabi splitting. If the system is not intensive, its properties depend on its size and the $\sqrt{N}$ factor from Eq.~\eqref{eq:dickeH} can be incorporated into definition of coupling $g$. This changes the critical coupling to $g_c =\sqrt{\omega \Omega/N}$ and the vacuum Rabi splitting to $\omega \pm g\sqrt{N}/2$ known from the cavity quantum electrodynamics experiments and Tavis-Cummings model. 
Therefore, the frequencies $\tilde \omega$ and $\tilde \Omega$ from Eq.~\eqref{eq:noentD} can be interpreted as the generalized vacuum Rabi splitting. Because of the weak coupling, even if it was possible to turn off the natural light-matter interaction, the amount of squeezing would be negligible and therefore virtually immeasurable. The number of virtual excitations as a function of $g/g_c$
\begin{align}
    \langle \hat a^\dagger \hat a \rangle + \langle \hat b^\dagger \hat b \rangle  = \langle \hat c^\dagger \hat c \rangle + \langle \hat d^\dagger \hat d \rangle = \sinh^2\xi_+ + \sinh^2\xi_-
\end{align}
is presented in Fig.~\ref{fig:fig2} and~\ref{fig:fig4}. As can be clearly seen, the light-matter coupling $g$ has to be extremely close to the critical coupling $g_c$ to create a significant amount of virtual excitations. This is one of the reasons why detecting them is so elusive. 

\section{finite size calculations}
\begin{figure}[htb!]
    \centering
    \includegraphics[width=0.49\textwidth]{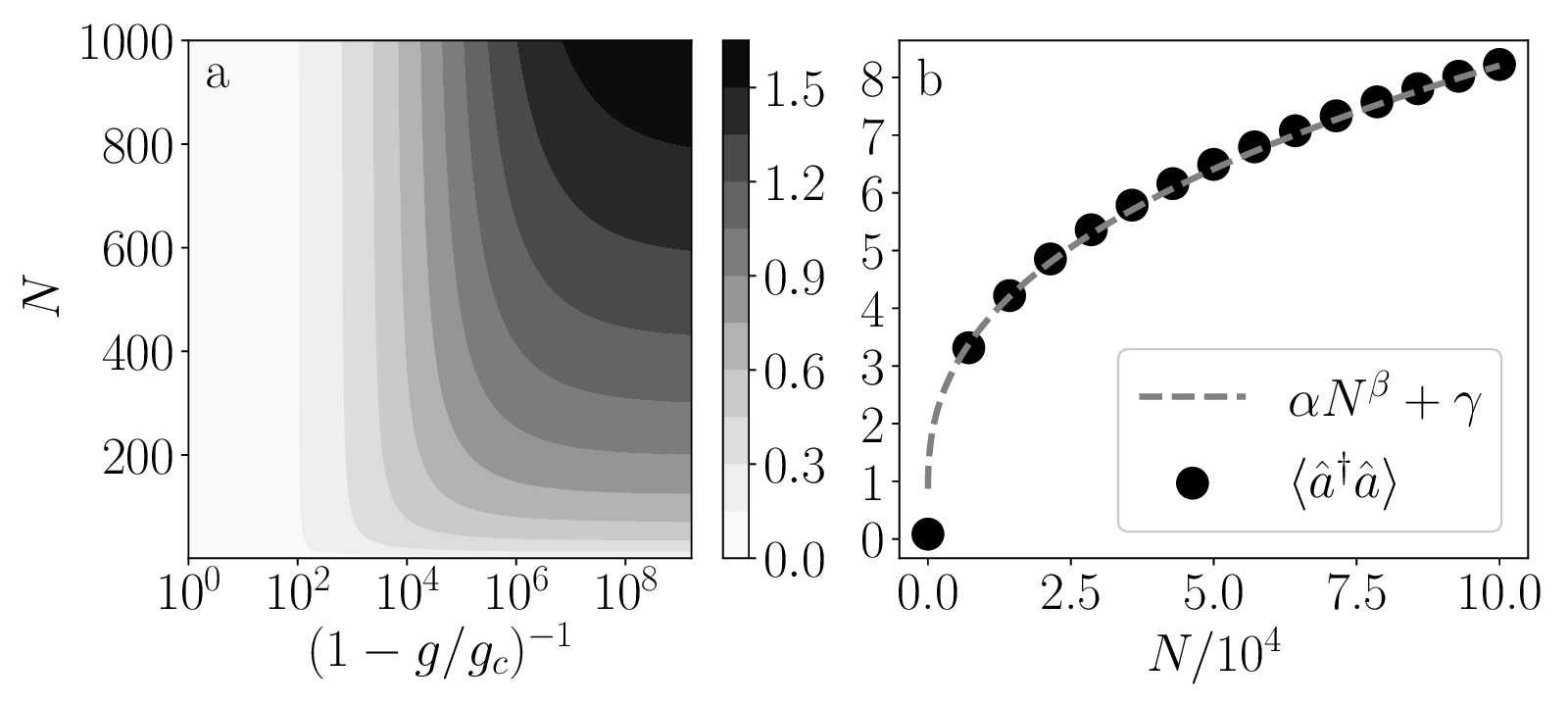}
    \caption{The number of virtual excitations (photons) as a function of the system size $N$ and the coupling strength $g/g_c$. (a) by increasing the number of spins $N$, the system can support more virtual excitations. (b) depicts scaling of the virtual excitations number at the critical point $g/g_c=1$ (black dots) and a fit (dashed gray line) with the leading order in power of $N$: $\langle \hat a^\dagger \hat a\rangle = \alpha N^\beta + \gamma$ with $\alpha \approx 13.5 $, $\beta \approx 0.4 $, and $\gamma \approx 0.8 $. Note that for the resonant case $\omega=\Omega$, the virtual excitations are equally squeezed $\langle \hat a^\dagger \hat a\rangle = \langle \hat b^\dagger \hat b\rangle $. }
    \label{fig:fig4}
\end{figure}

Previously, we have concentrated on the $N\rightarrow \infty$ limit of the Dicke model~\eqref{eq:dickeH} where the ground state can support infinite amount of virtual squeezing as a consequence of closing the energy gap. However, it is more important to know the maximal amount of virtual squeezing that can be generated in finite size systems, where the gap cannot be closed completely. To this end, we numerically diagonalize the Dicke model and calculate the number of virtual excitations (we look only at virtual photons $\langle \hat a^\dagger \hat a \rangle$) as a function of the system size $N$ and the coupling strength $g/g_c$. The results of numerical simulations are presented in Fig.~\ref{fig:fig4}. The number of virtual excitations grows very slowly with the system size which is another reason why the observation of virtual excitations remains so elusive in systems where only one two-level system is coupled to a mode of electromagnetic field. A numerical fit indicates that up to the leading order in $N$, the maximal number of virtual excitations can be expressed as $\langle \hat a^\dagger \hat a\rangle = \alpha N^\beta + \gamma$ with $\alpha \approx 13.5 $, $\beta \approx 0.4 $, and $\gamma \approx 0.8 $. 

\section{Conclusions}
In conclusion, by establishing a connection between squeezing, virtual excitations, and frequency shifts (closing and opening energy gap) of physical modes, we argue that the vacuum Rabi splitting is a manifestation of energetically protected virtual entanglement in the form of exotic two-mode squeezing at the onset of the superradiant phase transition. To this end, we analyze the Dicke model and its thermodynamic limit in the framework of bare and physical modes. We elucidate a straightforward connection between squeezing parameters of the bare modes, $\xi_\mp = \frac{1}{4}\log[1\mp g/g_c]$ and corresponding frequency shifts of the physical modes $\tilde \omega = \omega e^{2\xi_-}$ and $\tilde \Omega = \omega e^{2\xi_+}$. This suggests that squeezing is closely connected with closing energy gap in many-body quantum systems and measuring frequency shifts allows for reconstruction of the squeezing parameters. Since squeezing parameters are related to entanglement, the frequency shifts can be understood as a measure of entanglement between electromagnetic field vacuum and atoms. To the best of our knowledge, despite a vast literature related to squeezing and entanglement in the Dicke model~\cite{chaosDicke2003BrandesEmary,lambert2004entangleddicke,spinsqueezing2009song,garraway2011dicke,feshke2012qptdicke,vacuumsqueezing2017hu,universalfluc2020dickeshapiro}, these connections seem to be missing.

Restoring to numerical calculations, we find the limitation on the number of virtual excitations caused by the finite size of the system, and discuss why converting them into real ones is so elusive. The results of this work can be straightforwardly extended beyond the resonance case $\omega = \Omega$. In the case of $\omega \neq \Omega$, the squeezing of the virtual excitations will not be symmetric. For example, in the extreme case of $\omega/\Omega \rightarrow 0$ (or $\Omega/\omega \rightarrow 0$) only one harmonic oscillator experiences squeezing~\cite{chaosDicke2003BrandesEmary,plenio2015qrm} (see also Appendix for virtual squeezing in the quantum Rabi model). 

As in this manuscript we focus on the ground state physics, the future investigation will address how virtual squeezing picture can be applied to driven out-of-equilibrium systems where the effective Hamiltonian can be described by the Dicke model. In particular, in the driven-dissipative Dicke phase transition, realized by coupling external degrees of freedom of a Bose–Einstein condensate to the light field of a high-finesse optical cavity~\cite{ritsch2002selforg,DM2019domokos}, the electromagnetic radiation can be artificially coupled to matter and the coupling strength could be easily tuned over a broad range of values~\cite{esslinger2010srpt,esslinger2012rotontype,esslinger2013fluctuations,esslinger2015dynamicstructure}. This allows for investigating the physics close to the critical point of the superradiant phase transition where the energy gap almost closes and the squeezing of polaritons could be potentially observed by non-adiabatically varying the pump strength.

On a more general level, we would like to point out that what seems as frequency (or energy) shifts for physical modes might be related to squeezing of the virtual excitations. This suggests that properties of physical modes reflect the properties of virtual excitations. Therefore virtual squeezing of quantum fields, especially in the context of various quantum phase transitions~\cite{Sachdev_1999,Vojta_2003,Mazumdar_2019,gietka2022speedup} matched with their universality~\cite{sachdev_2011,zurek2014unive,plenio2015qrm}, might be a much more fundamental mechanism in strongly coupled light-matter systems and perhaps even in quantum field theory. 

\acknowledgements
K.G. acknowledges Arkadiusz Kosior, Laurin Ostermann, Christoph Hotter, Tobias Donner, and Helmut Ritsch for discussions. Simulations were performed using the open-source \textsc{QuantumOptics.jl}~\cite{kramer2018quantumoptics} framework in \textsc{Julia}. This work was supported by the Lise-Meitner Fellowship M3304-N of the Austrian Science Fund (FWF).

\section*{Appendix: Virtual squeezing in the quantum Rabi model}
\renewcommand\theequation{A\arabic{equation}}
\setcounter{equation}{0}
In the main text, we established a connection between virtual squeezing and frequency shifts in the Dicke model, which can be approximated by two interacting harmonic oscillators. A simpler system where a similar connection can be established is the quantum Rabi model in which there is only one two-level system coupled to a harmonic oscillator
\begin{align}
    \hat H = \omega \hat a^\dagger \hat a + \frac{\Omega}{2} \hat \sigma_z + \frac{g}{2}\left(\hat a + \hat a^\dagger \right)\hat \sigma_x.
\end{align}
In the limit of $\omega/\Omega \rightarrow 0$, quantum Rabi model can be approximated as only one harmonic mode which experiences squeezing as the system approaches the critical point of the superradiant phase transition
\begin{align}~\label{eq:qra}
    \hat H = \omega \hat a^\dagger \hat a - \frac{g^2}{4\Omega}\left( \hat a + \hat a^\dagger \right)^2  ,
\end{align}
whose eigenstates are squeezed number states
\begin{align}
    |n_\xi\rangle = \exp\bigg\{\frac{1}{2}\left(\xi^*\hat a^2-\xi\hat a^{\dagger2}\right)\bigg\}|n\rangle,
\end{align}
where $\xi = \frac{1}{4} \ln\{1-g^2/g_c^2\}$ is the squeezing parameter and $|n\rangle$ are number states generated by $\hat a^\dagger$ acting on the vacuum~$|0\rangle$. In the context of cavity quantum electrodynamics, naively interpreting $\hat a$ ($ \hat a^\dagger$) as an operator annihilating (creating) a photon of frequency $\omega$, one would come to the conclusion that the number of photons inside the cavity is
\begin{align}
    \langle \hat n \rangle  \equiv \langle \hat a^\dagger \hat a \rangle  = \sinh^2 \xi.
\end{align}
In fact, whenever $g\neq 0$, the operators $\hat a$ and $\hat a^\dagger$ do not correspond to physical modes. The physical modes can be found by diagonalizing the Hamiltonian from Eq.~\eqref{eq:qra}
\begin{align}
    \hat H = \tilde \omega \hat c^\dagger \hat c,
\end{align}
where $\hat c$ represents an excitation of the electromagnetic field with frequency $\tilde \omega = \omega \sqrt{1-g^2/g_c^2}= \omega e^{2\xi}$ related to the squeezing parameter $\xi$, and is defined as
\begin{align}
    \hat c = \cosh \xi \,\hat a + \sinh \xi\, \hat a^\dagger.
\end{align}
In the limit of $g\ll g_c$, the squeezing of the virtual excitations is negligible as $\cosh \xi \approx 1$ and $\sinh \xi \approx 0$, and the physical modes are almost identical to virtual modes $\hat a \approx \hat c$ which is the essence of the Jaynes-Cummings approximation.

Converting virtual excitations into real ones could be achieved by turning off the light-matter coupling instantaneously. This would effectively quench the frequency of the harmonic oscillator (electromagnetic field). As a result, the ground state of a strongly interacting system (very low frequency harmonic oscillator), becomes a highly excited state of the non-interacting system  (harmonic oscillator with frequency $\omega$). Since the system becomes suddenly excited, it can lose energy. Unfortunately, generating a significant amount of virtual excitations would require an extreme ratio $\omega/\Omega$ and an extraordinary control over the coupling parameter $g/g_c$, as presented in Fig.~\ref{fig:QRMs}.

\setcounter{figure}{0}
\renewcommand{\figurename}{Fig.}
\renewcommand{\thefigure}{A\arabic{figure}}

\begin{figure}[htb!]
    \centering
    \includegraphics[width=0.49\textwidth]{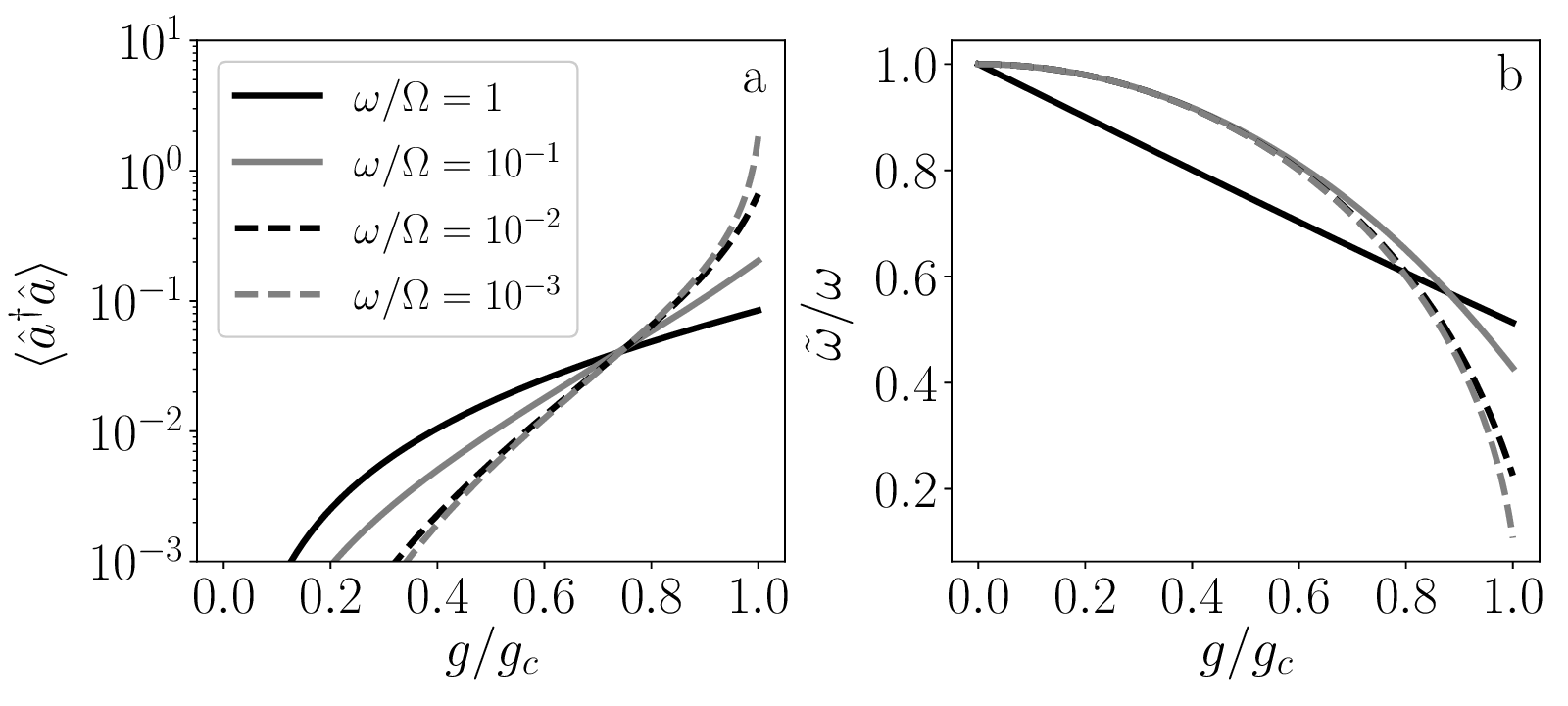}
    \caption{Virtual squeezing and its relation to frequency shifts (closing energy gap) in the quantum Rabi model as a function of $g/g_c$ and $\omega/\Omega$. (a) depicts the growth of virtual photons caused by squeezing of the ground state whereas (b) depicts corresponding frequency shifts of physical modes which can be associated with closing the energy gap at the onset of the superradiant phase transition.}
    \label{fig:QRMs}
\end{figure}


\begin{thebibliography}{75}%
\makeatletter
\providecommand \@ifxundefined [1]{%
 \@ifx{#1\undefined}
}%
\providecommand \@ifnum [1]{%
 \ifnum #1\expandafter \@firstoftwo
 \else \expandafter \@secondoftwo
 \fi
}%
\providecommand \@ifx [1]{%
 \ifx #1\expandafter \@firstoftwo
 \else \expandafter \@secondoftwo
 \fi
}%
\providecommand \natexlab [1]{#1}%
\providecommand \enquote  [1]{``#1''}%
\providecommand \bibnamefont  [1]{#1}%
\providecommand \bibfnamefont [1]{#1}%
\providecommand \citenamefont [1]{#1}%
\providecommand \href@noop [0]{\@secondoftwo}%
\providecommand \href [0]{\begingroup \@sanitize@url \@href}%
\providecommand \@href[1]{\@@startlink{#1}\@@href}%
\providecommand \@@href[1]{\endgroup#1\@@endlink}%
\providecommand \@sanitize@url [0]{\catcode `\\12\catcode `\$12\catcode `\&12\catcode `\#12\catcode `\^12\catcode `\_12\catcode `\%12\relax}%
\providecommand \@@startlink[1]{}%
\providecommand \@@endlink[0]{}%
\providecommand \url  [0]{\begingroup\@sanitize@url \@url }%
\providecommand \@url [1]{\endgroup\@href {#1}{\urlprefix }}%
\providecommand \urlprefix  [0]{URL }%
\providecommand \Eprint [0]{\href }%
\providecommand \doibase [0]{https://doi.org/}%
\providecommand \selectlanguage [0]{\@gobble}%
\providecommand \bibinfo  [0]{\@secondoftwo}%
\providecommand \bibfield  [0]{\@secondoftwo}%
\providecommand \translation [1]{[#1]}%
\providecommand \BibitemOpen [0]{}%
\providecommand \bibitemStop [0]{}%
\providecommand \bibitemNoStop [0]{.\EOS\space}%
\providecommand \EOS [0]{\spacefactor3000\relax}%
\providecommand \BibitemShut  [1]{\csname bibitem#1\endcsname}%
\let\auto@bib@innerbib\@empty
\bibitem [{\citenamefont {MacFarlane}\ \emph {et~al.}(2003)\citenamefont {MacFarlane}, \citenamefont {Dowling},\ and\ \citenamefont {Milburn}}]{QT2003milburn}%
  \BibitemOpen
  \bibfield  {author} {\bibinfo {author} {\bibfnamefont {A.~G.~J.}\ \bibnamefont {MacFarlane}}, \bibinfo {author} {\bibfnamefont {J.~P.}\ \bibnamefont {Dowling}},\ and\ \bibinfo {author} {\bibfnamefont {G.~J.}\ \bibnamefont {Milburn}},\ }\bibfield  {title} {\bibinfo {title} {Quantum technology: the second quantum revolution},\ }\href {https://doi.org/10.1098/rsta.2003.1227} {\bibfield  {journal} {\bibinfo  {journal} {Philos. Trans. Royal Soc. A}\ }\textbf {\bibinfo {volume} {361}},\ \bibinfo {pages} {1655} (\bibinfo {year} {2003})}\BibitemShut {NoStop}%
\bibitem [{\citenamefont {Acín}\ \emph {et~al.}(2018)\citenamefont {Acín}, \citenamefont {Bloch}, \citenamefont {Buhrman}, \citenamefont {Calarco}, \citenamefont {Eichler}, \citenamefont {Eisert}, \citenamefont {Esteve}, \citenamefont {Gisin}, \citenamefont {Glaser}, \citenamefont {Jelezko}, \citenamefont {Kuhr}, \citenamefont {Lewenstein}, \citenamefont {Riedel}, \citenamefont {Schmidt}, \citenamefont {Thew}, \citenamefont {Wallraff}, \citenamefont {Walmsley},\ and\ \citenamefont {Wilhelm}}]{Acín_2018}%
  \BibitemOpen
  \bibfield  {author} {\bibinfo {author} {\bibfnamefont {A.}~\bibnamefont {Acín}}, \bibinfo {author} {\bibfnamefont {I.}~\bibnamefont {Bloch}}, \bibinfo {author} {\bibfnamefont {H.}~\bibnamefont {Buhrman}}, \bibinfo {author} {\bibfnamefont {T.}~\bibnamefont {Calarco}}, \bibinfo {author} {\bibfnamefont {C.}~\bibnamefont {Eichler}}, \bibinfo {author} {\bibfnamefont {J.}~\bibnamefont {Eisert}}, \bibinfo {author} {\bibfnamefont {D.}~\bibnamefont {Esteve}}, \bibinfo {author} {\bibfnamefont {N.}~\bibnamefont {Gisin}}, \bibinfo {author} {\bibfnamefont {S.~J.}\ \bibnamefont {Glaser}}, \bibinfo {author} {\bibfnamefont {F.}~\bibnamefont {Jelezko}}, \bibinfo {author} {\bibfnamefont {S.}~\bibnamefont {Kuhr}}, \bibinfo {author} {\bibfnamefont {M.}~\bibnamefont {Lewenstein}}, \bibinfo {author} {\bibfnamefont {M.~F.}\ \bibnamefont {Riedel}}, \bibinfo {author} {\bibfnamefont {P.~O.}\ \bibnamefont {Schmidt}}, \bibinfo {author} {\bibfnamefont {R.}~\bibnamefont {Thew}}, \bibinfo {author} {\bibfnamefont {A.}~\bibnamefont
  {Wallraff}}, \bibinfo {author} {\bibfnamefont {I.}~\bibnamefont {Walmsley}},\ and\ \bibinfo {author} {\bibfnamefont {F.~K.}\ \bibnamefont {Wilhelm}},\ }\bibfield  {title} {\bibinfo {title} {The quantum technologies roadmap: a {E}uropean community view},\ }\href {https://doi.org/10.1088/1367-2630/aad1ea} {\bibfield  {journal} {\bibinfo  {journal} {New J. Phys.}\ }\textbf {\bibinfo {volume} {20}},\ \bibinfo {pages} {080201} (\bibinfo {year} {2018})}\BibitemShut {NoStop}%
\bibitem [{\citenamefont {Scully}\ and\ \citenamefont {Zubairy}(1999)}]{scully1999quantum}%
  \BibitemOpen
  \bibfield  {author} {\bibinfo {author} {\bibfnamefont {M.~O.}\ \bibnamefont {Scully}}\ and\ \bibinfo {author} {\bibfnamefont {M.~S.}\ \bibnamefont {Zubairy}},\ }\href {https://doi.org/https://doi.org/10.1017/CBO9780511813993} {\emph {\bibinfo {title} {Quantum {O}ptics}}}\ (\bibinfo  {publisher} {Cambridge University Press},\ \bibinfo {year} {1999})\BibitemShut {NoStop}%
\bibitem [{\citenamefont {Weiner}\ and\ \citenamefont {Ho}(2003)}]{weiner2003light}%
  \BibitemOpen
  \bibfield  {author} {\bibinfo {author} {\bibfnamefont {J.}~\bibnamefont {Weiner}}\ and\ \bibinfo {author} {\bibfnamefont {P.-T.}\ \bibnamefont {Ho}},\ }\href {https://books.google.at/books?id=JcufNIGv9eIC} {\emph {\bibinfo {title} {Light-{M}atter {I}nteraction: {F}undamentals and {A}pplications}}}\ (\bibinfo  {publisher} {Wiley},\ \bibinfo {year} {2003})\BibitemShut {NoStop}%
\bibitem [{\citenamefont {Walmsley}(2015)}]{walmsley2015scienceQO}%
  \BibitemOpen
  \bibfield  {author} {\bibinfo {author} {\bibfnamefont {I.~A.}\ \bibnamefont {Walmsley}},\ }\bibfield  {title} {\bibinfo {title} {Quantum optics: {S}cience and technology in a new light},\ }\href {https://doi.org/10.1126/science.aab0097} {\bibfield  {journal} {\bibinfo  {journal} {Science}\ }\textbf {\bibinfo {volume} {348}},\ \bibinfo {pages} {525} (\bibinfo {year} {2015})}\BibitemShut {NoStop}%
\bibitem [{\citenamefont {Greiner}\ and\ \citenamefont {Reinhardt}(2013)}]{greiner2013quantum}%
  \BibitemOpen
  \bibfield  {author} {\bibinfo {author} {\bibfnamefont {W.}~\bibnamefont {Greiner}}\ and\ \bibinfo {author} {\bibfnamefont {J.}~\bibnamefont {Reinhardt}},\ }\href {https://books.google.at/books?id=6y3rCAAAQBAJ} {\emph {\bibinfo {title} {Quantum {E}lectrodynamics}}}\ (\bibinfo  {publisher} {Springer Berlin Heidelberg},\ \bibinfo {year} {2013})\BibitemShut {NoStop}%
\bibitem [{\citenamefont {Lamb}\ and\ \citenamefont {Retherford}(1947)}]{lambshif1947}%
  \BibitemOpen
  \bibfield  {author} {\bibinfo {author} {\bibfnamefont {W.~E.}\ \bibnamefont {Lamb}}\ and\ \bibinfo {author} {\bibfnamefont {R.~C.}\ \bibnamefont {Retherford}},\ }\bibfield  {title} {\bibinfo {title} {Fine {S}tructure of the {H}ydrogen {A}tom by a {M}icrowave {M}ethod},\ }\href {https://doi.org/10.1103/PhysRev.72.241} {\bibfield  {journal} {\bibinfo  {journal} {Phys. Rev.}\ }\textbf {\bibinfo {volume} {72}},\ \bibinfo {pages} {241} (\bibinfo {year} {1947})}\BibitemShut {NoStop}%
\bibitem [{\citenamefont {Schwinger}(1948)}]{electronmoment1948schwinger}%
  \BibitemOpen
  \bibfield  {author} {\bibinfo {author} {\bibfnamefont {J.}~\bibnamefont {Schwinger}},\ }\bibfield  {title} {\bibinfo {title} {On quantum-electrodynamics and the magnetic moment of the electron},\ }\href {https://doi.org/10.1103/PhysRev.73.416} {\bibfield  {journal} {\bibinfo  {journal} {Phys. Rev.}\ }\textbf {\bibinfo {volume} {73}},\ \bibinfo {pages} {416} (\bibinfo {year} {1948})}\BibitemShut {NoStop}%
\bibitem [{\citenamefont {Cao}(2004)}]{cao2004conceptual}%
  \BibitemOpen
  \bibfield  {author} {\bibinfo {author} {\bibfnamefont {T.~Y.}\ \bibnamefont {Cao}},\ }\href@noop {} {\emph {\bibinfo {title} {Conceptual {F}oundations of {Q}uantum {F}ield {T}heory}}}\ (\bibinfo  {publisher} {Cambridge University Press},\ \bibinfo {year} {2004})\BibitemShut {NoStop}%
\bibitem [{\citenamefont {Ciuti}\ \emph {et~al.}(2005)\citenamefont {Ciuti}, \citenamefont {Bastard},\ and\ \citenamefont {Carusotto}}]{carusotto2005vacuum}%
  \BibitemOpen
  \bibfield  {author} {\bibinfo {author} {\bibfnamefont {C.}~\bibnamefont {Ciuti}}, \bibinfo {author} {\bibfnamefont {G.}~\bibnamefont {Bastard}},\ and\ \bibinfo {author} {\bibfnamefont {I.}~\bibnamefont {Carusotto}},\ }\bibfield  {title} {\bibinfo {title} {Quantum vacuum properties of the intersubband cavity polariton field},\ }\href {https://doi.org/10.1103/PhysRevB.72.115303} {\bibfield  {journal} {\bibinfo  {journal} {Phys. Rev. B}\ }\textbf {\bibinfo {volume} {72}},\ \bibinfo {pages} {115303} (\bibinfo {year} {2005})}\BibitemShut {NoStop}%
\bibitem [{\citenamefont {Loudon}(2000)}]{loudon2000quantum}%
  \BibitemOpen
  \bibfield  {author} {\bibinfo {author} {\bibfnamefont {R.}~\bibnamefont {Loudon}},\ }\href@noop {} {\emph {\bibinfo {title} {The {Q}uantum {T}heory of {L}ight}}}\ (\bibinfo  {publisher} {Oxford University Press},\ \bibinfo {year} {2000})\BibitemShut {NoStop}%
\bibitem [{\citenamefont {Forn-D\'{\i}az}\ \emph {et~al.}(2019)\citenamefont {Forn-D\'{\i}az}, \citenamefont {Lamata}, \citenamefont {Rico}, \citenamefont {Kono},\ and\ \citenamefont {Solano}}]{solano2019rmpUSC}%
  \BibitemOpen
  \bibfield  {author} {\bibinfo {author} {\bibfnamefont {P.}~\bibnamefont {Forn-D\'{\i}az}}, \bibinfo {author} {\bibfnamefont {L.}~\bibnamefont {Lamata}}, \bibinfo {author} {\bibfnamefont {E.}~\bibnamefont {Rico}}, \bibinfo {author} {\bibfnamefont {J.}~\bibnamefont {Kono}},\ and\ \bibinfo {author} {\bibfnamefont {E.}~\bibnamefont {Solano}},\ }\bibfield  {title} {\bibinfo {title} {Ultrastrong coupling regimes of light-matter interaction},\ }\href {https://doi.org/10.1103/RevModPhys.91.025005} {\bibfield  {journal} {\bibinfo  {journal} {Rev. Mod. Phys.}\ }\textbf {\bibinfo {volume} {91}},\ \bibinfo {pages} {025005} (\bibinfo {year} {2019})}\BibitemShut {NoStop}%
\bibitem [{\citenamefont {Frisk~Kockum}\ \emph {et~al.}(2019)\citenamefont {Frisk~Kockum}, \citenamefont {Miranowicz}, \citenamefont {De~Liberato}, \citenamefont {Savasta},\ and\ \citenamefont {Nori}}]{nori2019reviewUSC}%
  \BibitemOpen
  \bibfield  {author} {\bibinfo {author} {\bibfnamefont {A.}~\bibnamefont {Frisk~Kockum}}, \bibinfo {author} {\bibfnamefont {A.}~\bibnamefont {Miranowicz}}, \bibinfo {author} {\bibfnamefont {S.}~\bibnamefont {De~Liberato}}, \bibinfo {author} {\bibfnamefont {S.}~\bibnamefont {Savasta}},\ and\ \bibinfo {author} {\bibfnamefont {F.}~\bibnamefont {Nori}},\ }\bibfield  {title} {\bibinfo {title} {Ultrastrong coupling between light and matter},\ }\href {https://doi.org/10.1038/s42254-018-0006-2} {\bibfield  {journal} {\bibinfo  {journal} {Nat. Rev. Phys.}\ }\textbf {\bibinfo {volume} {1}},\ \bibinfo {pages} {19} (\bibinfo {year} {2019})}\BibitemShut {NoStop}%
\bibitem [{\citenamefont {Liberato}\ \emph {et~al.}(2007)\citenamefont {Liberato}, \citenamefont {Ciuti},\ and\ \citenamefont {Carusotto}}]{carusotto2007vacuumradiation}%
  \BibitemOpen
  \bibfield  {author} {\bibinfo {author} {\bibfnamefont {S.~D.}\ \bibnamefont {Liberato}}, \bibinfo {author} {\bibfnamefont {C.}~\bibnamefont {Ciuti}},\ and\ \bibinfo {author} {\bibfnamefont {I.}~\bibnamefont {Carusotto}},\ }\bibfield  {title} {\bibinfo {title} {Quantum {V}acuum {R}adiation {S}pectra from a {S}emiconductor {M}icrocavity with a {T}ime-{M}odulated {V}acuum {R}abi {F}requency},\ }\href {https://doi.org/10.1103/PhysRevLett.98.103602} {\bibfield  {journal} {\bibinfo  {journal} {Phys. Rev. Lett.}\ }\textbf {\bibinfo {volume} {98}},\ \bibinfo {pages} {103602} (\bibinfo {year} {2007})}\BibitemShut {NoStop}%
\bibitem [{\citenamefont {Werlang}\ \emph {et~al.}(2008)\citenamefont {Werlang}, \citenamefont {Dodonov}, \citenamefont {Duzzioni},\ and\ \citenamefont {Villas-B\^oas}}]{boas2008rabivacuum}%
  \BibitemOpen
  \bibfield  {author} {\bibinfo {author} {\bibfnamefont {T.}~\bibnamefont {Werlang}}, \bibinfo {author} {\bibfnamefont {A.~V.}\ \bibnamefont {Dodonov}}, \bibinfo {author} {\bibfnamefont {E.~I.}\ \bibnamefont {Duzzioni}},\ and\ \bibinfo {author} {\bibfnamefont {C.~J.}\ \bibnamefont {Villas-B\^oas}},\ }\bibfield  {title} {\bibinfo {title} {Rabi model beyond the rotating-wave approximation: {G}eneration of photons from vacuum through decoherence},\ }\href {https://doi.org/10.1103/PhysRevA.78.053805} {\bibfield  {journal} {\bibinfo  {journal} {Phys. Rev. A}\ }\textbf {\bibinfo {volume} {78}},\ \bibinfo {pages} {053805} (\bibinfo {year} {2008})}\BibitemShut {NoStop}%
\bibitem [{\citenamefont {Stassi}\ \emph {et~al.}(2013)\citenamefont {Stassi}, \citenamefont {Ridolfo}, \citenamefont {Di~Stefano}, \citenamefont {Hartmann},\ and\ \citenamefont {Savasta}}]{savasta2013virtualtoreal}%
  \BibitemOpen
  \bibfield  {author} {\bibinfo {author} {\bibfnamefont {R.}~\bibnamefont {Stassi}}, \bibinfo {author} {\bibfnamefont {A.}~\bibnamefont {Ridolfo}}, \bibinfo {author} {\bibfnamefont {O.}~\bibnamefont {Di~Stefano}}, \bibinfo {author} {\bibfnamefont {M.~J.}\ \bibnamefont {Hartmann}},\ and\ \bibinfo {author} {\bibfnamefont {S.}~\bibnamefont {Savasta}},\ }\bibfield  {title} {\bibinfo {title} {Spontaneous conversion from virtual to real photons in the ultrastrong-coupling regime},\ }\href {https://doi.org/10.1103/PhysRevLett.110.243601} {\bibfield  {journal} {\bibinfo  {journal} {Phys. Rev. Lett.}\ }\textbf {\bibinfo {volume} {110}},\ \bibinfo {pages} {243601} (\bibinfo {year} {2013})}\BibitemShut {NoStop}%
\bibitem [{\citenamefont {Cirio}\ \emph {et~al.}(2016)\citenamefont {Cirio}, \citenamefont {De~Liberato}, \citenamefont {Lambert},\ and\ \citenamefont {Nori}}]{nori2016electrolumin}%
  \BibitemOpen
  \bibfield  {author} {\bibinfo {author} {\bibfnamefont {M.}~\bibnamefont {Cirio}}, \bibinfo {author} {\bibfnamefont {S.}~\bibnamefont {De~Liberato}}, \bibinfo {author} {\bibfnamefont {N.}~\bibnamefont {Lambert}},\ and\ \bibinfo {author} {\bibfnamefont {F.}~\bibnamefont {Nori}},\ }\bibfield  {title} {\bibinfo {title} {Ground state electroluminescence},\ }\href {https://doi.org/10.1103/PhysRevLett.116.113601} {\bibfield  {journal} {\bibinfo  {journal} {Phys. Rev. Lett.}\ }\textbf {\bibinfo {volume} {116}},\ \bibinfo {pages} {113601} (\bibinfo {year} {2016})}\BibitemShut {NoStop}%
\bibitem [{\citenamefont {Cirio}\ \emph {et~al.}(2017)\citenamefont {Cirio}, \citenamefont {Debnath}, \citenamefont {Lambert},\ and\ \citenamefont {Nori}}]{nori2017virtualpressure}%
  \BibitemOpen
  \bibfield  {author} {\bibinfo {author} {\bibfnamefont {M.}~\bibnamefont {Cirio}}, \bibinfo {author} {\bibfnamefont {K.}~\bibnamefont {Debnath}}, \bibinfo {author} {\bibfnamefont {N.}~\bibnamefont {Lambert}},\ and\ \bibinfo {author} {\bibfnamefont {F.}~\bibnamefont {Nori}},\ }\bibfield  {title} {\bibinfo {title} {Amplified optomechanical transduction of virtual radiation pressure},\ }\href {https://doi.org/10.1103/PhysRevLett.119.053601} {\bibfield  {journal} {\bibinfo  {journal} {Phys. Rev. Lett.}\ }\textbf {\bibinfo {volume} {119}},\ \bibinfo {pages} {053601} (\bibinfo {year} {2017})}\BibitemShut {NoStop}%
\bibitem [{\citenamefont {Giannelli}\ \emph {et~al.}(2024)\citenamefont {Giannelli}, \citenamefont {Paladino}, \citenamefont {Grajcar}, \citenamefont {Paraoanu},\ and\ \citenamefont {Falci}}]{paraoanu2024detectvirtual}%
  \BibitemOpen
  \bibfield  {author} {\bibinfo {author} {\bibfnamefont {L.}~\bibnamefont {Giannelli}}, \bibinfo {author} {\bibfnamefont {E.}~\bibnamefont {Paladino}}, \bibinfo {author} {\bibfnamefont {M.}~\bibnamefont {Grajcar}}, \bibinfo {author} {\bibfnamefont {G.~S.}\ \bibnamefont {Paraoanu}},\ and\ \bibinfo {author} {\bibfnamefont {G.}~\bibnamefont {Falci}},\ }\bibfield  {title} {\bibinfo {title} {Detecting virtual photons in ultrastrongly coupled superconducting quantum circuits},\ }\href {https://doi.org/10.1103/PhysRevResearch.6.013008} {\bibfield  {journal} {\bibinfo  {journal} {Phys. Rev. Res.}\ }\textbf {\bibinfo {volume} {6}},\ \bibinfo {pages} {013008} (\bibinfo {year} {2024})}\BibitemShut {NoStop}%
\bibitem [{\citenamefont {Ritsch}\ \emph {et~al.}(2013)\citenamefont {Ritsch}, \citenamefont {Domokos}, \citenamefont {Brennecke},\ and\ \citenamefont {Esslinger}}]{helmut2013rmp}%
  \BibitemOpen
  \bibfield  {author} {\bibinfo {author} {\bibfnamefont {H.}~\bibnamefont {Ritsch}}, \bibinfo {author} {\bibfnamefont {P.}~\bibnamefont {Domokos}}, \bibinfo {author} {\bibfnamefont {F.}~\bibnamefont {Brennecke}},\ and\ \bibinfo {author} {\bibfnamefont {T.}~\bibnamefont {Esslinger}},\ }\bibfield  {title} {\bibinfo {title} {Cold atoms in cavity-generated dynamical optical potentials},\ }\href {https://doi.org/10.1103/RevModPhys.85.553} {\bibfield  {journal} {\bibinfo  {journal} {Rev. Mod. Phys.}\ }\textbf {\bibinfo {volume} {85}},\ \bibinfo {pages} {553} (\bibinfo {year} {2013})}\BibitemShut {NoStop}%
\bibitem [{\citenamefont {Rabi}(1937)}]{rabimodel1937}%
  \BibitemOpen
  \bibfield  {author} {\bibinfo {author} {\bibfnamefont {I.~I.}\ \bibnamefont {Rabi}},\ }\bibfield  {title} {\bibinfo {title} {Space {Q}uantization in a {G}yrating {M}agnetic {F}ield},\ }\href {https://doi.org/10.1103/PhysRev.51.652} {\bibfield  {journal} {\bibinfo  {journal} {Phys. Rev.}\ }\textbf {\bibinfo {volume} {51}},\ \bibinfo {pages} {652} (\bibinfo {year} {1937})}\BibitemShut {NoStop}%
\bibitem [{\citenamefont {Larson}\ and\ \citenamefont {Mavrogordatos}(2021)}]{JC2021larson}%
  \BibitemOpen
  \bibfield  {author} {\bibinfo {author} {\bibfnamefont {J.}~\bibnamefont {Larson}}\ and\ \bibinfo {author} {\bibfnamefont {T.}~\bibnamefont {Mavrogordatos}},\ }\href {https://doi.org/10.1088/978-0-7503-3447-1} {\emph {\bibinfo {title} {The {J}aynes–{C}ummings {M}odel and {I}ts {D}escendants}}},\ 2053-2563\ (\bibinfo  {publisher} {IOP Publishing},\ \bibinfo {year} {2021})\BibitemShut {NoStop}%
\bibitem [{\citenamefont {Sanchez-Mondragon}\ \emph {et~al.}(1983)\citenamefont {Sanchez-Mondragon}, \citenamefont {Narozhny},\ and\ \citenamefont {Eberly}}]{vrs2983eberly}%
  \BibitemOpen
  \bibfield  {author} {\bibinfo {author} {\bibfnamefont {J.~J.}\ \bibnamefont {Sanchez-Mondragon}}, \bibinfo {author} {\bibfnamefont {N.~B.}\ \bibnamefont {Narozhny}},\ and\ \bibinfo {author} {\bibfnamefont {J.~H.}\ \bibnamefont {Eberly}},\ }\bibfield  {title} {\bibinfo {title} {Theory of {S}pontaneous-{E}mission {L}ine {S}hape in an {I}deal {C}avity},\ }\href {https://doi.org/10.1103/PhysRevLett.51.550} {\bibfield  {journal} {\bibinfo  {journal} {Phys. Rev. Lett.}\ }\textbf {\bibinfo {volume} {51}},\ \bibinfo {pages} {550} (\bibinfo {year} {1983})}\BibitemShut {NoStop}%
\bibitem [{\citenamefont {Agarwal}(1984)}]{vrs2984agarwal}%
  \BibitemOpen
  \bibfield  {author} {\bibinfo {author} {\bibfnamefont {G.~S.}\ \bibnamefont {Agarwal}},\ }\bibfield  {title} {\bibinfo {title} {Vacuum-{F}ield {R}abi {S}plittings in {M}icrowave {A}bsorption by {R}ydberg {A}toms in a {C}avity},\ }\href {https://doi.org/10.1103/PhysRevLett.53.1732} {\bibfield  {journal} {\bibinfo  {journal} {Phys. Rev. Lett.}\ }\textbf {\bibinfo {volume} {53}},\ \bibinfo {pages} {1732} (\bibinfo {year} {1984})}\BibitemShut {NoStop}%
\bibitem [{\citenamefont {Thompson}\ \emph {et~al.}(1992)\citenamefont {Thompson}, \citenamefont {Rempe},\ and\ \citenamefont {Kimble}}]{vrs1992kimble}%
  \BibitemOpen
  \bibfield  {author} {\bibinfo {author} {\bibfnamefont {R.~J.}\ \bibnamefont {Thompson}}, \bibinfo {author} {\bibfnamefont {G.}~\bibnamefont {Rempe}},\ and\ \bibinfo {author} {\bibfnamefont {H.~J.}\ \bibnamefont {Kimble}},\ }\bibfield  {title} {\bibinfo {title} {Observation of normal-mode splitting for an atom in an optical cavity},\ }\href {https://doi.org/10.1103/PhysRevLett.68.1132} {\bibfield  {journal} {\bibinfo  {journal} {Phys. Rev. Lett.}\ }\textbf {\bibinfo {volume} {68}},\ \bibinfo {pages} {1132} (\bibinfo {year} {1992})}\BibitemShut {NoStop}%
\bibitem [{\citenamefont {Chen}\ \emph {et~al.}(2011)\citenamefont {Chen}, \citenamefont {Bohnet}, \citenamefont {Sankar}, \citenamefont {Dai},\ and\ \citenamefont {Thompson}}]{thompson2011vrsSS}%
  \BibitemOpen
  \bibfield  {author} {\bibinfo {author} {\bibfnamefont {Z.}~\bibnamefont {Chen}}, \bibinfo {author} {\bibfnamefont {J.~G.}\ \bibnamefont {Bohnet}}, \bibinfo {author} {\bibfnamefont {S.~R.}\ \bibnamefont {Sankar}}, \bibinfo {author} {\bibfnamefont {J.}~\bibnamefont {Dai}},\ and\ \bibinfo {author} {\bibfnamefont {J.~K.}\ \bibnamefont {Thompson}},\ }\bibfield  {title} {\bibinfo {title} {Conditional spin squeezing of a large ensemble via the vacuum rabi splitting},\ }\href {https://doi.org/10.1103/PhysRevLett.106.133601} {\bibfield  {journal} {\bibinfo  {journal} {Phys. Rev. Lett.}\ }\textbf {\bibinfo {volume} {106}},\ \bibinfo {pages} {133601} (\bibinfo {year} {2011})}\BibitemShut {NoStop}%
\bibitem [{\citenamefont {Ma}\ \emph {et~al.}(2011)\citenamefont {Ma}, \citenamefont {Wang}, \citenamefont {Sun},\ and\ \citenamefont {Nori}}]{MA201189}%
  \BibitemOpen
  \bibfield  {author} {\bibinfo {author} {\bibfnamefont {J.}~\bibnamefont {Ma}}, \bibinfo {author} {\bibfnamefont {X.}~\bibnamefont {Wang}}, \bibinfo {author} {\bibfnamefont {C.}~\bibnamefont {Sun}},\ and\ \bibinfo {author} {\bibfnamefont {F.}~\bibnamefont {Nori}},\ }\bibfield  {title} {\bibinfo {title} {Quantum spin squeezing},\ }\href {https://doi.org/https://doi.org/10.1016/j.physrep.2011.08.003} {\bibfield  {journal} {\bibinfo  {journal} {Phys. Rep.}\ }\textbf {\bibinfo {volume} {509}},\ \bibinfo {pages} {89} (\bibinfo {year} {2011})}\BibitemShut {NoStop}%
\bibitem [{\citenamefont {Weisbuch}\ \emph {et~al.}(1992)\citenamefont {Weisbuch}, \citenamefont {Nishioka}, \citenamefont {Ishikawa},\ and\ \citenamefont {Arakawa}}]{vrs1992ssmc}%
  \BibitemOpen
  \bibfield  {author} {\bibinfo {author} {\bibfnamefont {C.}~\bibnamefont {Weisbuch}}, \bibinfo {author} {\bibfnamefont {M.}~\bibnamefont {Nishioka}}, \bibinfo {author} {\bibfnamefont {A.}~\bibnamefont {Ishikawa}},\ and\ \bibinfo {author} {\bibfnamefont {Y.}~\bibnamefont {Arakawa}},\ }\bibfield  {title} {\bibinfo {title} {Observation of the coupled exciton-photon mode splitting in a semiconductor quantum microcavity},\ }\href {https://doi.org/10.1103/PhysRevLett.69.3314} {\bibfield  {journal} {\bibinfo  {journal} {Phys. Rev. Lett.}\ }\textbf {\bibinfo {volume} {69}},\ \bibinfo {pages} {3314} (\bibinfo {year} {1992})}\BibitemShut {NoStop}%
\bibitem [{\citenamefont {G{\"u}nter}\ \emph {et~al.}(2009)\citenamefont {G{\"u}nter}, \citenamefont {Anappara}, \citenamefont {Hees}, \citenamefont {Sell}, \citenamefont {Biasiol}, \citenamefont {Sorba}, \citenamefont {De~Liberato}, \citenamefont {Ciuti}, \citenamefont {Tredicucci}, \citenamefont {Leitenstorfer},\ and\ \citenamefont {Huber}}]{ciurri2009nature}%
  \BibitemOpen
  \bibfield  {author} {\bibinfo {author} {\bibfnamefont {G.}~\bibnamefont {G{\"u}nter}}, \bibinfo {author} {\bibfnamefont {A.~A.}\ \bibnamefont {Anappara}}, \bibinfo {author} {\bibfnamefont {J.}~\bibnamefont {Hees}}, \bibinfo {author} {\bibfnamefont {A.}~\bibnamefont {Sell}}, \bibinfo {author} {\bibfnamefont {G.}~\bibnamefont {Biasiol}}, \bibinfo {author} {\bibfnamefont {L.}~\bibnamefont {Sorba}}, \bibinfo {author} {\bibfnamefont {S.}~\bibnamefont {De~Liberato}}, \bibinfo {author} {\bibfnamefont {C.}~\bibnamefont {Ciuti}}, \bibinfo {author} {\bibfnamefont {A.}~\bibnamefont {Tredicucci}}, \bibinfo {author} {\bibfnamefont {A.}~\bibnamefont {Leitenstorfer}},\ and\ \bibinfo {author} {\bibfnamefont {R.}~\bibnamefont {Huber}},\ }\bibfield  {title} {\bibinfo {title} {Sub-cycle switch-on of ultrastrong light--matter interaction},\ }\href {https://doi.org/10.1038/nature07838} {\bibfield  {journal} {\bibinfo  {journal} {Nature}\ }\textbf {\bibinfo {volume} {458}},\ \bibinfo {pages} {178} (\bibinfo {year}
  {2009})}\BibitemShut {NoStop}%
\bibitem [{\citenamefont {Graham}(1987)}]{squeezing1987frequency}%
  \BibitemOpen
  \bibfield  {author} {\bibinfo {author} {\bibfnamefont {R.}~\bibnamefont {Graham}},\ }\bibfield  {title} {\bibinfo {title} {Squeezing and frequency changes in harmonic oscillations},\ }\href {https://doi.org/10.1080/09500348714550801} {\bibfield  {journal} {\bibinfo  {journal} {J. Mod. Opt.}\ }\textbf {\bibinfo {volume} {34}},\ \bibinfo {pages} {873} (\bibinfo {year} {1987})}\BibitemShut {NoStop}%
\bibitem [{\citenamefont {Lo}(1990)}]{CFLo_1990_squeezingfrequency}%
  \BibitemOpen
  \bibfield  {author} {\bibinfo {author} {\bibfnamefont {C.~F.}\ \bibnamefont {Lo}},\ }\bibfield  {title} {\bibinfo {title} {Squeezing by tuning the oscillator frequency},\ }\href {https://doi.org/10.1088/0305-4470/23/7/021} {\bibfield  {journal} {\bibinfo  {journal} {J. Phys. A Math. Gen.}\ }\textbf {\bibinfo {volume} {23}},\ \bibinfo {pages} {1155} (\bibinfo {year} {1990})}\BibitemShut {NoStop}%
\bibitem [{\citenamefont {Garraway}(2011{\natexlab{a}})}]{dickemodel2011garraway}%
  \BibitemOpen
  \bibfield  {author} {\bibinfo {author} {\bibfnamefont {B.~M.}\ \bibnamefont {Garraway}},\ }\bibfield  {title} {\bibinfo {title} {The {D}icke model in quantum optics: {D}icke model revisited},\ }\href {https://doi.org/10.1098/rsta.2010.0333} {\bibfield  {journal} {\bibinfo  {journal} {Philos. Trans. R. Soc. A}\ }\textbf {\bibinfo {volume} {369}},\ \bibinfo {pages} {1137} (\bibinfo {year} {2011}{\natexlab{a}})}\BibitemShut {NoStop}%
\bibitem [{\citenamefont {Hepp}\ and\ \citenamefont {Lieb}(1973)}]{HEPP1973360}%
  \BibitemOpen
  \bibfield  {author} {\bibinfo {author} {\bibfnamefont {K.}~\bibnamefont {Hepp}}\ and\ \bibinfo {author} {\bibfnamefont {E.~H.}\ \bibnamefont {Lieb}},\ }\bibfield  {title} {\bibinfo {title} {On the superradiant phase transition for molecules in a quantized radiation field: the {D}icke maser model},\ }\href {https://doi.org/https://doi.org/10.1016/0003-4916(73)90039-0} {\bibfield  {journal} {\bibinfo  {journal} {Ann. Phys.}\ }\textbf {\bibinfo {volume} {76}},\ \bibinfo {pages} {360} (\bibinfo {year} {1973})}\BibitemShut {NoStop}%
\bibitem [{\citenamefont {Wang}\ and\ \citenamefont {Hioe}(1973)}]{srpt1973hioe}%
  \BibitemOpen
  \bibfield  {author} {\bibinfo {author} {\bibfnamefont {Y.~K.}\ \bibnamefont {Wang}}\ and\ \bibinfo {author} {\bibfnamefont {F.~T.}\ \bibnamefont {Hioe}},\ }\bibfield  {title} {\bibinfo {title} {Phase {T}ransition in the {D}icke {M}odel of {S}uperradiance},\ }\href {https://doi.org/10.1103/PhysRevA.7.831} {\bibfield  {journal} {\bibinfo  {journal} {Phys. Rev. A}\ }\textbf {\bibinfo {volume} {7}},\ \bibinfo {pages} {831} (\bibinfo {year} {1973})}\BibitemShut {NoStop}%
\bibitem [{\citenamefont {Rza\ifmmode~\dot{z}\else \.{z}\fi{}ewski}\ \emph {et~al.}(1975)\citenamefont {Rza\ifmmode~\dot{z}\else \.{z}\fi{}ewski}, \citenamefont {W\'odkiewicz},\ and\ \citenamefont {\ifmmode~\dot{Z}\else \.{Z}\fi{}akowicz}}]{wodkiewicz1975ptnogo}%
  \BibitemOpen
  \bibfield  {author} {\bibinfo {author} {\bibfnamefont {K.}~\bibnamefont {Rza\ifmmode~\dot{z}\else \.{z}\fi{}ewski}}, \bibinfo {author} {\bibfnamefont {K.}~\bibnamefont {W\'odkiewicz}},\ and\ \bibinfo {author} {\bibfnamefont {W.}~\bibnamefont {\ifmmode~\dot{Z}\else \.{Z}\fi{}akowicz}},\ }\bibfield  {title} {\bibinfo {title} {Phase {T}ransitions, {T}wo-{L}evel {A}toms, and the ${A}^{2}$ {T}erm},\ }\href {https://doi.org/10.1103/PhysRevLett.35.432} {\bibfield  {journal} {\bibinfo  {journal} {Phys. Rev. Lett.}\ }\textbf {\bibinfo {volume} {35}},\ \bibinfo {pages} {432} (\bibinfo {year} {1975})}\BibitemShut {NoStop}%
\bibitem [{\citenamefont {Bialynicki-Birula}\ and\ \citenamefont {Rza\ifmmode \mbox{\c{}}\else \c{}\fi{}\ifmmode~\dot{z}\else \.{z}\fi{}ewski}(1979)}]{birula1979nogoSR}%
  \BibitemOpen
  \bibfield  {author} {\bibinfo {author} {\bibfnamefont {I.}~\bibnamefont {Bialynicki-Birula}}\ and\ \bibinfo {author} {\bibfnamefont {K.}~\bibnamefont {Rza\ifmmode \mbox{\c{}}\else \c{}\fi{}\ifmmode~\dot{z}\else \.{z}\fi{}ewski}},\ }\bibfield  {title} {\bibinfo {title} {No-go theorem concerning the superradiant phase transition in atomic systems},\ }\href {https://doi.org/10.1103/PhysRevA.19.301} {\bibfield  {journal} {\bibinfo  {journal} {Phys. Rev. A}\ }\textbf {\bibinfo {volume} {19}},\ \bibinfo {pages} {301} (\bibinfo {year} {1979})}\BibitemShut {NoStop}%
\bibitem [{\citenamefont {Emary}\ and\ \citenamefont {Brandes}(2003)}]{chaosDicke2003BrandesEmary}%
  \BibitemOpen
  \bibfield  {author} {\bibinfo {author} {\bibfnamefont {C.}~\bibnamefont {Emary}}\ and\ \bibinfo {author} {\bibfnamefont {T.}~\bibnamefont {Brandes}},\ }\bibfield  {title} {\bibinfo {title} {Chaos and the quantum phase transition in the {D}icke model},\ }\href {https://doi.org/10.1103/PhysRevE.67.066203} {\bibfield  {journal} {\bibinfo  {journal} {Phys. Rev. E}\ }\textbf {\bibinfo {volume} {67}},\ \bibinfo {pages} {066203} (\bibinfo {year} {2003})}\BibitemShut {NoStop}%
\bibitem [{\citenamefont {Maschler}\ \emph {et~al.}(2007)\citenamefont {Maschler}, \citenamefont {Ritsch}, \citenamefont {Vukics},\ and\ \citenamefont {Domokos}}]{MASCHLER2007446}%
  \BibitemOpen
  \bibfield  {author} {\bibinfo {author} {\bibfnamefont {C.}~\bibnamefont {Maschler}}, \bibinfo {author} {\bibfnamefont {H.}~\bibnamefont {Ritsch}}, \bibinfo {author} {\bibfnamefont {A.}~\bibnamefont {Vukics}},\ and\ \bibinfo {author} {\bibfnamefont {P.}~\bibnamefont {Domokos}},\ }\bibfield  {title} {\bibinfo {title} {Entanglement assisted fast reordering of atoms in an optical lattice within a cavity at {T}=0},\ }\href {https://doi.org/https://doi.org/10.1016/j.optcom.2007.01.069} {\bibfield  {journal} {\bibinfo  {journal} {Opt. Commun.}\ }\textbf {\bibinfo {volume} {273}},\ \bibinfo {pages} {446} (\bibinfo {year} {2007})}\BibitemShut {NoStop}%
\bibitem [{\citenamefont {Viehmann}\ \emph {et~al.}(2011)\citenamefont {Viehmann}, \citenamefont {von Delft},\ and\ \citenamefont {Marquardt}}]{SRPT2011marquardt}%
  \BibitemOpen
  \bibfield  {author} {\bibinfo {author} {\bibfnamefont {O.}~\bibnamefont {Viehmann}}, \bibinfo {author} {\bibfnamefont {J.}~\bibnamefont {von Delft}},\ and\ \bibinfo {author} {\bibfnamefont {F.}~\bibnamefont {Marquardt}},\ }\bibfield  {title} {\bibinfo {title} {Superradiant {P}hase {T}ransitions and the {S}tandard {D}escription of {C}ircuit {Q}{E}{D}},\ }\href {https://doi.org/10.1103/PhysRevLett.107.113602} {\bibfield  {journal} {\bibinfo  {journal} {Phys. Rev. Lett.}\ }\textbf {\bibinfo {volume} {107}},\ \bibinfo {pages} {113602} (\bibinfo {year} {2011})}\BibitemShut {NoStop}%
\bibitem [{\citenamefont {Bakemeier}\ \emph {et~al.}(2012)\citenamefont {Bakemeier}, \citenamefont {Alvermann},\ and\ \citenamefont {Fehske}}]{feshke2012qptdicke}%
  \BibitemOpen
  \bibfield  {author} {\bibinfo {author} {\bibfnamefont {L.}~\bibnamefont {Bakemeier}}, \bibinfo {author} {\bibfnamefont {A.}~\bibnamefont {Alvermann}},\ and\ \bibinfo {author} {\bibfnamefont {H.}~\bibnamefont {Fehske}},\ }\bibfield  {title} {\bibinfo {title} {Quantum phase transition in the {D}icke model with critical and noncritical entanglement},\ }\href {https://doi.org/10.1103/PhysRevA.85.043821} {\bibfield  {journal} {\bibinfo  {journal} {Phys. Rev. A}\ }\textbf {\bibinfo {volume} {85}},\ \bibinfo {pages} {043821} (\bibinfo {year} {2012})}\BibitemShut {NoStop}%
\bibitem [{\citenamefont {Hwang}\ \emph {et~al.}(2015)\citenamefont {Hwang}, \citenamefont {Puebla},\ and\ \citenamefont {Plenio}}]{plenio2015qrm}%
  \BibitemOpen
  \bibfield  {author} {\bibinfo {author} {\bibfnamefont {M.-J.}\ \bibnamefont {Hwang}}, \bibinfo {author} {\bibfnamefont {R.}~\bibnamefont {Puebla}},\ and\ \bibinfo {author} {\bibfnamefont {M.~B.}\ \bibnamefont {Plenio}},\ }\bibfield  {title} {\bibinfo {title} {Quantum {P}hase {T}ransition and {U}niversal {D}ynamics in the {R}abi {M}odel},\ }\href {https://doi.org/10.1103/PhysRevLett.115.180404} {\bibfield  {journal} {\bibinfo  {journal} {Phys. Rev. Lett.}\ }\textbf {\bibinfo {volume} {115}},\ \bibinfo {pages} {180404} (\bibinfo {year} {2015})}\BibitemShut {NoStop}%
\bibitem [{\citenamefont {Klinder}\ \emph {et~al.}(2015)\citenamefont {Klinder}, \citenamefont {Keßler}, \citenamefont {Wolke}, \citenamefont {Mathey},\ and\ \citenamefont {Hemmerich}}]{hemmerich2015dynamic}%
  \BibitemOpen
  \bibfield  {author} {\bibinfo {author} {\bibfnamefont {J.}~\bibnamefont {Klinder}}, \bibinfo {author} {\bibfnamefont {H.}~\bibnamefont {Keßler}}, \bibinfo {author} {\bibfnamefont {M.}~\bibnamefont {Wolke}}, \bibinfo {author} {\bibfnamefont {L.}~\bibnamefont {Mathey}},\ and\ \bibinfo {author} {\bibfnamefont {A.}~\bibnamefont {Hemmerich}},\ }\bibfield  {title} {\bibinfo {title} {Dynamical phase transition in the open {D}icke model},\ }\href {https://doi.org/10.1073/pnas.1417132112} {\bibfield  {journal} {\bibinfo  {journal} {PNAS}\ }\textbf {\bibinfo {volume} {112}},\ \bibinfo {pages} {3290} (\bibinfo {year} {2015})}\BibitemShut {NoStop}%
\bibitem [{\citenamefont {Larson}\ and\ \citenamefont {Irish}(2017)}]{Larson_2017}%
  \BibitemOpen
  \bibfield  {author} {\bibinfo {author} {\bibfnamefont {J.}~\bibnamefont {Larson}}\ and\ \bibinfo {author} {\bibfnamefont {E.~K.}\ \bibnamefont {Irish}},\ }\bibfield  {title} {\bibinfo {title} {Some remarks on ‘superradiant’ phase transitions in light-matter systems},\ }\href {https://doi.org/10.1088/1751-8121/aa65dc} {\bibfield  {journal} {\bibinfo  {journal} {J. Phys. A Math. Theor.}\ }\textbf {\bibinfo {volume} {50}},\ \bibinfo {pages} {174002} (\bibinfo {year} {2017})}\BibitemShut {NoStop}%
\bibitem [{\citenamefont {Safavi-Naini}\ \emph {et~al.}(2018)\citenamefont {Safavi-Naini}, \citenamefont {Lewis-Swan}, \citenamefont {Bohnet}, \citenamefont {G\"arttner}, \citenamefont {Gilmore}, \citenamefont {Jordan}, \citenamefont {Cohn}, \citenamefont {Freericks}, \citenamefont {Rey},\ and\ \citenamefont {Bollinger}}]{amr2018ionDM}%
  \BibitemOpen
  \bibfield  {author} {\bibinfo {author} {\bibfnamefont {A.}~\bibnamefont {Safavi-Naini}}, \bibinfo {author} {\bibfnamefont {R.~J.}\ \bibnamefont {Lewis-Swan}}, \bibinfo {author} {\bibfnamefont {J.~G.}\ \bibnamefont {Bohnet}}, \bibinfo {author} {\bibfnamefont {M.}~\bibnamefont {G\"arttner}}, \bibinfo {author} {\bibfnamefont {K.~A.}\ \bibnamefont {Gilmore}}, \bibinfo {author} {\bibfnamefont {J.~E.}\ \bibnamefont {Jordan}}, \bibinfo {author} {\bibfnamefont {J.}~\bibnamefont {Cohn}}, \bibinfo {author} {\bibfnamefont {J.~K.}\ \bibnamefont {Freericks}}, \bibinfo {author} {\bibfnamefont {A.~M.}\ \bibnamefont {Rey}},\ and\ \bibinfo {author} {\bibfnamefont {J.~J.}\ \bibnamefont {Bollinger}},\ }\bibfield  {title} {\bibinfo {title} {Verification of a {M}any-{I}on {S}imulator of the {D}icke {M}odel {T}hrough {S}low {Q}uenches across a {P}hase {T}ransition},\ }\href {https://doi.org/10.1103/PhysRevLett.121.040503} {\bibfield  {journal} {\bibinfo  {journal} {Phys. Rev. Lett.}\ }\textbf {\bibinfo {volume} {121}},\ \bibinfo
  {pages} {040503} (\bibinfo {year} {2018})}\BibitemShut {NoStop}%
\bibitem [{\citenamefont {Zheng}\ \emph {et~al.}(2023)\citenamefont {Zheng}, \citenamefont {Ning}, \citenamefont {Chen}, \citenamefont {L\"u}, \citenamefont {Shen}, \citenamefont {Xu}, \citenamefont {Zhang}, \citenamefont {Xu}, \citenamefont {Li}, \citenamefont {Xia}, \citenamefont {Wu}, \citenamefont {Yang}, \citenamefont {Miranowicz}, \citenamefont {Lambert}, \citenamefont {Zheng}, \citenamefont {Fan}, \citenamefont {Nori},\ and\ \citenamefont {Zheng}}]{nori2023srpt}%
  \BibitemOpen
  \bibfield  {author} {\bibinfo {author} {\bibfnamefont {R.-H.}\ \bibnamefont {Zheng}}, \bibinfo {author} {\bibfnamefont {W.}~\bibnamefont {Ning}}, \bibinfo {author} {\bibfnamefont {Y.-H.}\ \bibnamefont {Chen}}, \bibinfo {author} {\bibfnamefont {J.-H.}\ \bibnamefont {L\"u}}, \bibinfo {author} {\bibfnamefont {L.-T.}\ \bibnamefont {Shen}}, \bibinfo {author} {\bibfnamefont {K.}~\bibnamefont {Xu}}, \bibinfo {author} {\bibfnamefont {Y.-R.}\ \bibnamefont {Zhang}}, \bibinfo {author} {\bibfnamefont {D.}~\bibnamefont {Xu}}, \bibinfo {author} {\bibfnamefont {H.}~\bibnamefont {Li}}, \bibinfo {author} {\bibfnamefont {Y.}~\bibnamefont {Xia}}, \bibinfo {author} {\bibfnamefont {F.}~\bibnamefont {Wu}}, \bibinfo {author} {\bibfnamefont {Z.-B.}\ \bibnamefont {Yang}}, \bibinfo {author} {\bibfnamefont {A.}~\bibnamefont {Miranowicz}}, \bibinfo {author} {\bibfnamefont {N.}~\bibnamefont {Lambert}}, \bibinfo {author} {\bibfnamefont {D.}~\bibnamefont {Zheng}}, \bibinfo {author} {\bibfnamefont {H.}~\bibnamefont {Fan}}, \bibinfo
  {author} {\bibfnamefont {F.}~\bibnamefont {Nori}},\ and\ \bibinfo {author} {\bibfnamefont {S.-B.}\ \bibnamefont {Zheng}},\ }\bibfield  {title} {\bibinfo {title} {Observation of a {S}uperradiant {P}hase {T}ransition with {E}mergent {C}at {S}tates},\ }\href {https://doi.org/10.1103/PhysRevLett.131.113601} {\bibfield  {journal} {\bibinfo  {journal} {Phys. Rev. Lett.}\ }\textbf {\bibinfo {volume} {131}},\ \bibinfo {pages} {113601} (\bibinfo {year} {2023})}\BibitemShut {NoStop}%
\bibitem [{\citenamefont {Kim}\ \emph {et~al.}(2024)\citenamefont {Kim}, \citenamefont {Dasgupta}, \citenamefont {Ma}, \citenamefont {Park}, \citenamefont {Wei}, \citenamefont {Luo}, \citenamefont {Doumani}, \citenamefont {Li}, \citenamefont {Yang}, \citenamefont {Cheng}, \citenamefont {Kim}, \citenamefont {Everitt}, \citenamefont {Kimura}, \citenamefont {Nojiri}, \citenamefont {Wang}, \citenamefont {Cao}, \citenamefont {Bamba}, \citenamefont {Hazzard},\ and\ \citenamefont {Kono}}]{kim2024observation}%
  \BibitemOpen
  \bibfield  {author} {\bibinfo {author} {\bibfnamefont {D.}~\bibnamefont {Kim}}, \bibinfo {author} {\bibfnamefont {S.}~\bibnamefont {Dasgupta}}, \bibinfo {author} {\bibfnamefont {X.}~\bibnamefont {Ma}}, \bibinfo {author} {\bibfnamefont {J.-M.}\ \bibnamefont {Park}}, \bibinfo {author} {\bibfnamefont {H.-T.}\ \bibnamefont {Wei}}, \bibinfo {author} {\bibfnamefont {L.}~\bibnamefont {Luo}}, \bibinfo {author} {\bibfnamefont {J.}~\bibnamefont {Doumani}}, \bibinfo {author} {\bibfnamefont {X.}~\bibnamefont {Li}}, \bibinfo {author} {\bibfnamefont {W.}~\bibnamefont {Yang}}, \bibinfo {author} {\bibfnamefont {D.}~\bibnamefont {Cheng}}, \bibinfo {author} {\bibfnamefont {R.~H.~J.}\ \bibnamefont {Kim}}, \bibinfo {author} {\bibfnamefont {H.~O.}\ \bibnamefont {Everitt}}, \bibinfo {author} {\bibfnamefont {S.}~\bibnamefont {Kimura}}, \bibinfo {author} {\bibfnamefont {H.}~\bibnamefont {Nojiri}}, \bibinfo {author} {\bibfnamefont {J.}~\bibnamefont {Wang}}, \bibinfo {author} {\bibfnamefont {S.}~\bibnamefont {Cao}}, \bibinfo
  {author} {\bibfnamefont {M.}~\bibnamefont {Bamba}}, \bibinfo {author} {\bibfnamefont {K.~R.~A.}\ \bibnamefont {Hazzard}},\ and\ \bibinfo {author} {\bibfnamefont {J.}~\bibnamefont {Kono}},\ }\href@noop {} {\bibinfo {title} {Observation of the {M}agnonic {D}icke {S}uperradiant {P}hase {T}ransition}} (\bibinfo {year} {2024}),\ \Eprint {https://arxiv.org/abs/2401.01873} {arXiv:2401.01873 [quant-ph]} \BibitemShut {NoStop}%
\bibitem [{\citenamefont {Mivehvar}(2024)}]{farokh2024dickemodels}%
  \BibitemOpen
  \bibfield  {author} {\bibinfo {author} {\bibfnamefont {F.}~\bibnamefont {Mivehvar}},\ }\bibfield  {title} {\bibinfo {title} {Conventional and unconventional dicke models: Multistabilities and nonequilibrium dynamics},\ }\href {https://doi.org/10.1103/PhysRevLett.132.073602} {\bibfield  {journal} {\bibinfo  {journal} {Phys. Rev. Lett.}\ }\textbf {\bibinfo {volume} {132}},\ \bibinfo {pages} {073602} (\bibinfo {year} {2024})}\BibitemShut {NoStop}%
\bibitem [{\citenamefont {Baumann}\ \emph {et~al.}(2010)\citenamefont {Baumann}, \citenamefont {Guerlin}, \citenamefont {Brennecke},\ and\ \citenamefont {Esslinger}}]{esslinger2010srpt}%
  \BibitemOpen
  \bibfield  {author} {\bibinfo {author} {\bibfnamefont {K.}~\bibnamefont {Baumann}}, \bibinfo {author} {\bibfnamefont {C.}~\bibnamefont {Guerlin}}, \bibinfo {author} {\bibfnamefont {F.}~\bibnamefont {Brennecke}},\ and\ \bibinfo {author} {\bibfnamefont {T.}~\bibnamefont {Esslinger}},\ }\bibfield  {title} {\bibinfo {title} {Dicke quantum phase transition with a superfluid gas in an optical cavity},\ }\href {https://doi.org/10.1038/nature09009} {\bibfield  {journal} {\bibinfo  {journal} {Nature}\ }\textbf {\bibinfo {volume} {464}},\ \bibinfo {pages} {1301} (\bibinfo {year} {2010})}\BibitemShut {NoStop}%
\bibitem [{\citenamefont {Holstein}\ and\ \citenamefont {Primakoff}(1940)}]{HP1940PR}%
  \BibitemOpen
  \bibfield  {author} {\bibinfo {author} {\bibfnamefont {T.}~\bibnamefont {Holstein}}\ and\ \bibinfo {author} {\bibfnamefont {H.}~\bibnamefont {Primakoff}},\ }\bibfield  {title} {\bibinfo {title} {Field {D}ependence of the {I}ntrinsic {D}omain {M}agnetization of a {F}erromagnet},\ }\href {https://doi.org/10.1103/PhysRev.58.1098} {\bibfield  {journal} {\bibinfo  {journal} {Phys. Rev.}\ }\textbf {\bibinfo {volume} {58}},\ \bibinfo {pages} {1098} (\bibinfo {year} {1940})}\BibitemShut {NoStop}%
\bibitem [{\citenamefont {Schumaker}\ and\ \citenamefont {Caves}(1985)}]{twomodeS1985theory}%
  \BibitemOpen
  \bibfield  {author} {\bibinfo {author} {\bibfnamefont {B.~L.}\ \bibnamefont {Schumaker}}\ and\ \bibinfo {author} {\bibfnamefont {C.~M.}\ \bibnamefont {Caves}},\ }\bibfield  {title} {\bibinfo {title} {New formalism for two-photon quantum optics. {I}{I}. {M}athematical foundation and compact notation},\ }\href {https://doi.org/10.1103/PhysRevA.31.3093} {\bibfield  {journal} {\bibinfo  {journal} {Phys. Rev. A}\ }\textbf {\bibinfo {volume} {31}},\ \bibinfo {pages} {3093} (\bibinfo {year} {1985})}\BibitemShut {NoStop}%
\bibitem [{\citenamefont {Heidmann}\ \emph {et~al.}(1987)\citenamefont {Heidmann}, \citenamefont {Horowicz}, \citenamefont {Reynaud}, \citenamefont {Giacobino}, \citenamefont {Fabre},\ and\ \citenamefont {Camy}}]{twomodeS1987exp}%
  \BibitemOpen
  \bibfield  {author} {\bibinfo {author} {\bibfnamefont {A.}~\bibnamefont {Heidmann}}, \bibinfo {author} {\bibfnamefont {R.~J.}\ \bibnamefont {Horowicz}}, \bibinfo {author} {\bibfnamefont {S.}~\bibnamefont {Reynaud}}, \bibinfo {author} {\bibfnamefont {E.}~\bibnamefont {Giacobino}}, \bibinfo {author} {\bibfnamefont {C.}~\bibnamefont {Fabre}},\ and\ \bibinfo {author} {\bibfnamefont {G.}~\bibnamefont {Camy}},\ }\bibfield  {title} {\bibinfo {title} {Observation of {Q}uantum {N}oise {R}eduction on {T}win {L}aser {B}eams},\ }\href {https://doi.org/10.1103/PhysRevLett.59.2555} {\bibfield  {journal} {\bibinfo  {journal} {Phys. Rev. Lett.}\ }\textbf {\bibinfo {volume} {59}},\ \bibinfo {pages} {2555} (\bibinfo {year} {1987})}\BibitemShut {NoStop}%
\bibitem [{\citenamefont {Ciuti}\ and\ \citenamefont {Carusotto}(2006)}]{ciuti2006input}%
  \BibitemOpen
  \bibfield  {author} {\bibinfo {author} {\bibfnamefont {C.}~\bibnamefont {Ciuti}}\ and\ \bibinfo {author} {\bibfnamefont {I.}~\bibnamefont {Carusotto}},\ }\bibfield  {title} {\bibinfo {title} {Input-output theory of cavities in the ultrastrong coupling regime: {T}he case of time-independent cavity parameters},\ }\href {https://doi.org/10.1103/PhysRevA.74.033811} {\bibfield  {journal} {\bibinfo  {journal} {Phys. Rev. A}\ }\textbf {\bibinfo {volume} {74}},\ \bibinfo {pages} {033811} (\bibinfo {year} {2006})}\BibitemShut {NoStop}%
\bibitem [{\citenamefont {Basov}\ \emph {et~al.}(2021)\citenamefont {Basov}, \citenamefont {Asenjo-Garcia}, \citenamefont {Schuck}, \citenamefont {Zhu},\ and\ \citenamefont {Rubio}}]{Basov2021}%
  \BibitemOpen
  \bibfield  {author} {\bibinfo {author} {\bibfnamefont {D.~N.}\ \bibnamefont {Basov}}, \bibinfo {author} {\bibfnamefont {A.}~\bibnamefont {Asenjo-Garcia}}, \bibinfo {author} {\bibfnamefont {P.~J.}\ \bibnamefont {Schuck}}, \bibinfo {author} {\bibfnamefont {X.}~\bibnamefont {Zhu}},\ and\ \bibinfo {author} {\bibfnamefont {A.}~\bibnamefont {Rubio}},\ }\bibfield  {title} {\bibinfo {title} {Polariton panorama},\ }\href {https://doi.org/doi:10.1515/nanoph-2020-0449} {\bibfield  {journal} {\bibinfo  {journal} {Nanophotonics}\ }\textbf {\bibinfo {volume} {10}},\ \bibinfo {pages} {549} (\bibinfo {year} {2021})}\BibitemShut {NoStop}%
\bibitem [{\citenamefont {Farokh~Mivehvar}\ and\ \citenamefont {Ritsch}(2021)}]{novel2021QED}%
  \BibitemOpen
  \bibfield  {author} {\bibinfo {author} {\bibfnamefont {T.~D.}\ \bibnamefont {Farokh~Mivehvar}, \bibfnamefont {Francesco~Piazza}}\ and\ \bibinfo {author} {\bibfnamefont {H.}~\bibnamefont {Ritsch}},\ }\bibfield  {title} {\bibinfo {title} {Cavity qed with quantum gases: new paradigms in many-body physics},\ }\href {https://doi.org/10.1080/00018732.2021.1969727} {\bibfield  {journal} {\bibinfo  {journal} {Advances in Physics}\ }\textbf {\bibinfo {volume} {70}},\ \bibinfo {pages} {1} (\bibinfo {year} {2021})}\BibitemShut {NoStop}%
\bibitem [{\citenamefont {Beaudoin}\ \emph {et~al.}(2011)\citenamefont {Beaudoin}, \citenamefont {Gambetta},\ and\ \citenamefont {Blais}}]{blais2011dissipation}%
  \BibitemOpen
  \bibfield  {author} {\bibinfo {author} {\bibfnamefont {F.}~\bibnamefont {Beaudoin}}, \bibinfo {author} {\bibfnamefont {J.~M.}\ \bibnamefont {Gambetta}},\ and\ \bibinfo {author} {\bibfnamefont {A.}~\bibnamefont {Blais}},\ }\bibfield  {title} {\bibinfo {title} {Dissipation and ultrastrong coupling in circuit {Q}{E}{D}},\ }\href {https://doi.org/10.1103/PhysRevA.84.043832} {\bibfield  {journal} {\bibinfo  {journal} {Phys. Rev. A}\ }\textbf {\bibinfo {volume} {84}},\ \bibinfo {pages} {043832} (\bibinfo {year} {2011})}\BibitemShut {NoStop}%
\bibitem [{\citenamefont {Hwang}\ \emph {et~al.}(2018)\citenamefont {Hwang}, \citenamefont {Rabl},\ and\ \citenamefont {Plenio}}]{plenio2018dissprabi}%
  \BibitemOpen
  \bibfield  {author} {\bibinfo {author} {\bibfnamefont {M.-J.}\ \bibnamefont {Hwang}}, \bibinfo {author} {\bibfnamefont {P.}~\bibnamefont {Rabl}},\ and\ \bibinfo {author} {\bibfnamefont {M.~B.}\ \bibnamefont {Plenio}},\ }\bibfield  {title} {\bibinfo {title} {Dissipative phase transition in the open quantum {R}abi model},\ }\href {https://doi.org/10.1103/PhysRevA.97.013825} {\bibfield  {journal} {\bibinfo  {journal} {Phys. Rev. A}\ }\textbf {\bibinfo {volume} {97}},\ \bibinfo {pages} {013825} (\bibinfo {year} {2018})}\BibitemShut {NoStop}%
\bibitem [{\citenamefont {Cattaneo}\ \emph {et~al.}(2019)\citenamefont {Cattaneo}, \citenamefont {Giorgi}, \citenamefont {Maniscalco},\ and\ \citenamefont {Zambrini}}]{Cattaneo_2019}%
  \BibitemOpen
  \bibfield  {author} {\bibinfo {author} {\bibfnamefont {M.}~\bibnamefont {Cattaneo}}, \bibinfo {author} {\bibfnamefont {G.~L.}\ \bibnamefont {Giorgi}}, \bibinfo {author} {\bibfnamefont {S.}~\bibnamefont {Maniscalco}},\ and\ \bibinfo {author} {\bibfnamefont {R.}~\bibnamefont {Zambrini}},\ }\bibfield  {title} {\bibinfo {title} {Local versus global master equation with common and separate baths: superiority of the global approach in partial secular approximation},\ }\href {https://doi.org/10.1088/1367-2630/ab54ac} {\bibfield  {journal} {\bibinfo  {journal} {New J. Phys.}\ }\textbf {\bibinfo {volume} {21}},\ \bibinfo {pages} {113045} (\bibinfo {year} {2019})}\BibitemShut {NoStop}%
\bibitem [{\citenamefont {Gietka}\ \emph {et~al.}(2023)\citenamefont {Gietka}, \citenamefont {Hotter},\ and\ \citenamefont {Ritsch}}]{gietka2023uniqueSSqrm}%
  \BibitemOpen
  \bibfield  {author} {\bibinfo {author} {\bibfnamefont {K.}~\bibnamefont {Gietka}}, \bibinfo {author} {\bibfnamefont {C.}~\bibnamefont {Hotter}},\ and\ \bibinfo {author} {\bibfnamefont {H.}~\bibnamefont {Ritsch}},\ }\bibfield  {title} {\bibinfo {title} {Unique steady-state squeezing in a driven quantum rabi model},\ }\href {https://doi.org/10.1103/PhysRevLett.131.223604} {\bibfield  {journal} {\bibinfo  {journal} {Phys. Rev. Lett.}\ }\textbf {\bibinfo {volume} {131}},\ \bibinfo {pages} {223604} (\bibinfo {year} {2023})}\BibitemShut {NoStop}%
\bibitem [{\citenamefont {Lambert}\ \emph {et~al.}(2004)\citenamefont {Lambert}, \citenamefont {Emary},\ and\ \citenamefont {Brandes}}]{lambert2004entangleddicke}%
  \BibitemOpen
  \bibfield  {author} {\bibinfo {author} {\bibfnamefont {N.}~\bibnamefont {Lambert}}, \bibinfo {author} {\bibfnamefont {C.}~\bibnamefont {Emary}},\ and\ \bibinfo {author} {\bibfnamefont {T.}~\bibnamefont {Brandes}},\ }\bibfield  {title} {\bibinfo {title} {Entanglement and the phase transition in single-mode superradiance},\ }\href {https://doi.org/10.1103/PhysRevLett.92.073602} {\bibfield  {journal} {\bibinfo  {journal} {Phys. Rev. Lett.}\ }\textbf {\bibinfo {volume} {92}},\ \bibinfo {pages} {073602} (\bibinfo {year} {2004})}\BibitemShut {NoStop}%
\bibitem [{\citenamefont {Song}\ \emph {et~al.}(2009)\citenamefont {Song}, \citenamefont {Yan}, \citenamefont {Ma},\ and\ \citenamefont {Wang}}]{spinsqueezing2009song}%
  \BibitemOpen
  \bibfield  {author} {\bibinfo {author} {\bibfnamefont {L.}~\bibnamefont {Song}}, \bibinfo {author} {\bibfnamefont {D.}~\bibnamefont {Yan}}, \bibinfo {author} {\bibfnamefont {J.}~\bibnamefont {Ma}},\ and\ \bibinfo {author} {\bibfnamefont {X.}~\bibnamefont {Wang}},\ }\bibfield  {title} {\bibinfo {title} {Spin squeezing as an indicator of quantum chaos in the dicke model},\ }\href {https://doi.org/10.1103/PhysRevE.79.046220} {\bibfield  {journal} {\bibinfo  {journal} {Phys. Rev. E}\ }\textbf {\bibinfo {volume} {79}},\ \bibinfo {pages} {046220} (\bibinfo {year} {2009})}\BibitemShut {NoStop}%
\bibitem [{\citenamefont {Garraway}(2011{\natexlab{b}})}]{garraway2011dicke}%
  \BibitemOpen
  \bibfield  {author} {\bibinfo {author} {\bibfnamefont {B.~M.}\ \bibnamefont {Garraway}},\ }\bibfield  {title} {\bibinfo {title} {The dicke model in quantum optics: Dicke model revisited},\ }\href {https://doi.org/10.1098/rsta.2010.0333} {\bibfield  {journal} {\bibinfo  {journal} {Philos. Trans. R. Soc. A}\ }\textbf {\bibinfo {volume} {369}},\ \bibinfo {pages} {1137} (\bibinfo {year} {2011}{\natexlab{b}})}\BibitemShut {NoStop}%
\bibitem [{\citenamefont {Hu}\ \emph {et~al.}(2017)\citenamefont {Hu}, \citenamefont {Chen}, \citenamefont {Vendeiro}, \citenamefont {Urvoy}, \citenamefont {Braverman},\ and\ \citenamefont {Vuleti\ifmmode~\acute{c}\else \'{c}\fi{}}}]{vacuumsqueezing2017hu}%
  \BibitemOpen
  \bibfield  {author} {\bibinfo {author} {\bibfnamefont {J.}~\bibnamefont {Hu}}, \bibinfo {author} {\bibfnamefont {W.}~\bibnamefont {Chen}}, \bibinfo {author} {\bibfnamefont {Z.}~\bibnamefont {Vendeiro}}, \bibinfo {author} {\bibfnamefont {A.}~\bibnamefont {Urvoy}}, \bibinfo {author} {\bibfnamefont {B.}~\bibnamefont {Braverman}},\ and\ \bibinfo {author} {\bibfnamefont {V.}~\bibnamefont {Vuleti\ifmmode~\acute{c}\else \'{c}\fi{}}},\ }\bibfield  {title} {\bibinfo {title} {Vacuum spin squeezing},\ }\href {https://doi.org/10.1103/PhysRevA.96.050301} {\bibfield  {journal} {\bibinfo  {journal} {Phys. Rev. A}\ }\textbf {\bibinfo {volume} {96}},\ \bibinfo {pages} {050301} (\bibinfo {year} {2017})}\BibitemShut {NoStop}%
\bibitem [{\citenamefont {Shapiro}\ \emph {et~al.}(2020)\citenamefont {Shapiro}, \citenamefont {Pogosov},\ and\ \citenamefont {Lozovik}}]{universalfluc2020dickeshapiro}%
  \BibitemOpen
  \bibfield  {author} {\bibinfo {author} {\bibfnamefont {D.~S.}\ \bibnamefont {Shapiro}}, \bibinfo {author} {\bibfnamefont {W.~V.}\ \bibnamefont {Pogosov}},\ and\ \bibinfo {author} {\bibfnamefont {Y.~E.}\ \bibnamefont {Lozovik}},\ }\bibfield  {title} {\bibinfo {title} {Universal fluctuations and squeezing in a generalized dicke model near the superradiant phase transition},\ }\href {https://doi.org/10.1103/PhysRevA.102.023703} {\bibfield  {journal} {\bibinfo  {journal} {Phys. Rev. A}\ }\textbf {\bibinfo {volume} {102}},\ \bibinfo {pages} {023703} (\bibinfo {year} {2020})}\BibitemShut {NoStop}%
\bibitem [{\citenamefont {Domokos}\ and\ \citenamefont {Ritsch}(2002)}]{ritsch2002selforg}%
  \BibitemOpen
  \bibfield  {author} {\bibinfo {author} {\bibfnamefont {P.}~\bibnamefont {Domokos}}\ and\ \bibinfo {author} {\bibfnamefont {H.}~\bibnamefont {Ritsch}},\ }\bibfield  {title} {\bibinfo {title} {Collective {C}ooling and {S}elf-{O}rganization of {A}toms in a {C}avity},\ }\href {https://doi.org/10.1103/PhysRevLett.89.253003} {\bibfield  {journal} {\bibinfo  {journal} {Phys. Rev. Lett.}\ }\textbf {\bibinfo {volume} {89}},\ \bibinfo {pages} {253003} (\bibinfo {year} {2002})}\BibitemShut {NoStop}%
\bibitem [{\citenamefont {Nagy}\ \emph {et~al.}(2010)\citenamefont {Nagy}, \citenamefont {K\'onya}, \citenamefont {Szirmai},\ and\ \citenamefont {Domokos}}]{DM2019domokos}%
  \BibitemOpen
  \bibfield  {author} {\bibinfo {author} {\bibfnamefont {D.}~\bibnamefont {Nagy}}, \bibinfo {author} {\bibfnamefont {G.}~\bibnamefont {K\'onya}}, \bibinfo {author} {\bibfnamefont {G.}~\bibnamefont {Szirmai}},\ and\ \bibinfo {author} {\bibfnamefont {P.}~\bibnamefont {Domokos}},\ }\bibfield  {title} {\bibinfo {title} {Dicke-{M}odel {P}hase {T}ransition in the {Q}uantum {M}otion of a {B}ose-{E}instein {C}ondensate in an {O}ptical {C}avity},\ }\href {https://doi.org/10.1103/PhysRevLett.104.130401} {\bibfield  {journal} {\bibinfo  {journal} {Phys. Rev. Lett.}\ }\textbf {\bibinfo {volume} {104}},\ \bibinfo {pages} {130401} (\bibinfo {year} {2010})}\BibitemShut {NoStop}%
\bibitem [{\citenamefont {Mottl}\ \emph {et~al.}(2012)\citenamefont {Mottl}, \citenamefont {Brennecke}, \citenamefont {Baumann}, \citenamefont {Landig}, \citenamefont {Donner},\ and\ \citenamefont {Esslinger}}]{esslinger2012rotontype}%
  \BibitemOpen
  \bibfield  {author} {\bibinfo {author} {\bibfnamefont {R.}~\bibnamefont {Mottl}}, \bibinfo {author} {\bibfnamefont {F.}~\bibnamefont {Brennecke}}, \bibinfo {author} {\bibfnamefont {K.}~\bibnamefont {Baumann}}, \bibinfo {author} {\bibfnamefont {R.}~\bibnamefont {Landig}}, \bibinfo {author} {\bibfnamefont {T.}~\bibnamefont {Donner}},\ and\ \bibinfo {author} {\bibfnamefont {T.}~\bibnamefont {Esslinger}},\ }\bibfield  {title} {\bibinfo {title} {Roton-{T}ype {M}ode {S}oftening in a {Q}uantum {G}as with {C}avity-{M}ediated {L}ong-{R}ange {I}nteractions},\ }\href {https://doi.org/10.1126/science.1220314} {\bibfield  {journal} {\bibinfo  {journal} {Science}\ }\textbf {\bibinfo {volume} {336}},\ \bibinfo {pages} {1570} (\bibinfo {year} {2012})}\BibitemShut {NoStop}%
\bibitem [{\citenamefont {Brennecke}\ \emph {et~al.}(2013)\citenamefont {Brennecke}, \citenamefont {Mottl}, \citenamefont {Baumann}, \citenamefont {Landig}, \citenamefont {Donner},\ and\ \citenamefont {Esslinger}}]{esslinger2013fluctuations}%
  \BibitemOpen
  \bibfield  {author} {\bibinfo {author} {\bibfnamefont {F.}~\bibnamefont {Brennecke}}, \bibinfo {author} {\bibfnamefont {R.}~\bibnamefont {Mottl}}, \bibinfo {author} {\bibfnamefont {K.}~\bibnamefont {Baumann}}, \bibinfo {author} {\bibfnamefont {R.}~\bibnamefont {Landig}}, \bibinfo {author} {\bibfnamefont {T.}~\bibnamefont {Donner}},\ and\ \bibinfo {author} {\bibfnamefont {T.}~\bibnamefont {Esslinger}},\ }\bibfield  {title} {\bibinfo {title} {Real-time observation of fluctuations at the driven-dissipative dicke phase transition},\ }\href {https://doi.org/10.1073/pnas.1306993110} {\bibfield  {journal} {\bibinfo  {journal} {PNAS}\ }\textbf {\bibinfo {volume} {110}},\ \bibinfo {pages} {11763} (\bibinfo {year} {2013})}\BibitemShut {NoStop}%
\bibitem [{\citenamefont {Landig}\ \emph {et~al.}(2015)\citenamefont {Landig}, \citenamefont {Brennecke}, \citenamefont {Mottl}, \citenamefont {Donner},\ and\ \citenamefont {Esslinger}}]{esslinger2015dynamicstructure}%
  \BibitemOpen
  \bibfield  {author} {\bibinfo {author} {\bibfnamefont {R.}~\bibnamefont {Landig}}, \bibinfo {author} {\bibfnamefont {F.}~\bibnamefont {Brennecke}}, \bibinfo {author} {\bibfnamefont {R.}~\bibnamefont {Mottl}}, \bibinfo {author} {\bibfnamefont {T.}~\bibnamefont {Donner}},\ and\ \bibinfo {author} {\bibfnamefont {T.}~\bibnamefont {Esslinger}},\ }\bibfield  {title} {\bibinfo {title} {Measuring the dynamic structure factor of a quantum gas undergoing a structural phase transition},\ }\href {https://doi.org/10.1038/ncomms8046} {\bibfield  {journal} {\bibinfo  {journal} {Nat Commun}\ }\textbf {\bibinfo {volume} {6}},\ \bibinfo {pages} {7046} (\bibinfo {year} {2015})}\BibitemShut {NoStop}%
\bibitem [{\citenamefont {Sachdev}(1999)}]{Sachdev_1999}%
  \BibitemOpen
  \bibfield  {author} {\bibinfo {author} {\bibfnamefont {S.}~\bibnamefont {Sachdev}},\ }\bibfield  {title} {\bibinfo {title} {Quantum phase transitions},\ }\href {https://doi.org/10.1088/2058-7058/12/4/23} {\bibfield  {journal} {\bibinfo  {journal} {Phys. World}\ }\textbf {\bibinfo {volume} {12}},\ \bibinfo {pages} {33} (\bibinfo {year} {1999})}\BibitemShut {NoStop}%
\bibitem [{\citenamefont {Vojta}(2003)}]{Vojta_2003}%
  \BibitemOpen
  \bibfield  {author} {\bibinfo {author} {\bibfnamefont {M.}~\bibnamefont {Vojta}},\ }\bibfield  {title} {\bibinfo {title} {Quantum phase transitions},\ }\href {https://doi.org/10.1088/0034-4885/66/12/R01} {\bibfield  {journal} {\bibinfo  {journal} {Rep. Prog. Phys.}\ }\textbf {\bibinfo {volume} {66}},\ \bibinfo {pages} {2069} (\bibinfo {year} {2003})}\BibitemShut {NoStop}%
\bibitem [{\citenamefont {Mazumdar}\ and\ \citenamefont {White}(2019)}]{Mazumdar_2019}%
  \BibitemOpen
  \bibfield  {author} {\bibinfo {author} {\bibfnamefont {A.}~\bibnamefont {Mazumdar}}\ and\ \bibinfo {author} {\bibfnamefont {G.}~\bibnamefont {White}},\ }\bibfield  {title} {\bibinfo {title} {Review of cosmic phase transitions: their significance and experimental signatures},\ }\href {https://doi.org/10.1088/1361-6633/ab1f55} {\bibfield  {journal} {\bibinfo  {journal} {Rep. Prog. Phys.}\ }\textbf {\bibinfo {volume} {82}},\ \bibinfo {pages} {076901} (\bibinfo {year} {2019})}\BibitemShut {NoStop}%
\bibitem [{\citenamefont {Gietka}(2022)}]{gietka2022speedup}%
  \BibitemOpen
  \bibfield  {author} {\bibinfo {author} {\bibfnamefont {K.}~\bibnamefont {Gietka}},\ }\bibfield  {title} {\bibinfo {title} {Squeezing by critical speeding up: Applications in quantum metrology},\ }\href {https://doi.org/10.1103/PhysRevA.105.042620} {\bibfield  {journal} {\bibinfo  {journal} {Phys. Rev. A}\ }\textbf {\bibinfo {volume} {105}},\ \bibinfo {pages} {042620} (\bibinfo {year} {2022})}\BibitemShut {NoStop}%
\bibitem [{\citenamefont {Sachdev}(2011)}]{sachdev_2011}%
  \BibitemOpen
  \bibfield  {author} {\bibinfo {author} {\bibfnamefont {S.}~\bibnamefont {Sachdev}},\ }\href {https://doi.org/10.1017/CBO9780511973765} {\emph {\bibinfo {title} {Quantum {P}hase {T}ransitions}}},\ \bibinfo {edition} {2nd}\ ed.\ (\bibinfo  {publisher} {Cambridge University Press},\ \bibinfo {year} {2011})\BibitemShut {NoStop}%
\bibitem [{\citenamefont {del Campo}\ and\ \citenamefont {Zurek}(2014)}]{zurek2014unive}%
  \BibitemOpen
  \bibfield  {author} {\bibinfo {author} {\bibfnamefont {A.}~\bibnamefont {del Campo}}\ and\ \bibinfo {author} {\bibfnamefont {W.~H.}\ \bibnamefont {Zurek}},\ }\bibfield  {title} {\bibinfo {title} {Universality of phase transition dynamics: {T}opological defects from symmetry breaking},\ }\href {https://doi.org/10.1142/S0217751X1430018X} {\bibfield  {journal} {\bibinfo  {journal} {Int. J. Mod. Phys. A}\ }\textbf {\bibinfo {volume} {29}},\ \bibinfo {pages} {1430018} (\bibinfo {year} {2014})}\BibitemShut {NoStop}%
\bibitem [{\citenamefont {Krämer}\ \emph {et~al.}(2018)\citenamefont {Krämer}, \citenamefont {Plankensteiner}, \citenamefont {Ostermann},\ and\ \citenamefont {Ritsch}}]{kramer2018quantumoptics}%
  \BibitemOpen
  \bibfield  {author} {\bibinfo {author} {\bibfnamefont {S.}~\bibnamefont {Krämer}}, \bibinfo {author} {\bibfnamefont {D.}~\bibnamefont {Plankensteiner}}, \bibinfo {author} {\bibfnamefont {L.}~\bibnamefont {Ostermann}},\ and\ \bibinfo {author} {\bibfnamefont {H.}~\bibnamefont {Ritsch}},\ }\bibfield  {title} {\bibinfo {title} {Quantumoptics.jl: A {J}ulia framework for simulating open quantum systems},\ }\href {https://doi.org/https://doi.org/10.1016/j.cpc.2018.02.004} {\bibfield  {journal} {\bibinfo  {journal} {Comput. Phys. Commun.}\ }\textbf {\bibinfo {volume} {227}},\ \bibinfo {pages} {109} (\bibinfo {year} {2018})}\BibitemShut {NoStop}%
\end{thebibliography}
\end{document}